\documentclass[%
 reprint,
 amsmath,amssymb,
 aps,
 pra,
]{revtex4-2}
\usepackage[dvipsnames]{xcolor}
\usepackage[pdftex]{graphicx}
\usepackage{dcolumn}
\usepackage{bm}
\usepackage{soul}

\usepackage[colorlinks=true,linkcolor=blue,urlcolor=blue,citecolor=blue]{hyperref}
\usepackage[utf8]{inputenc}
\usepackage[T1]{fontenc}
\usepackage{braket}
\usepackage{physics}
\usepackage{tikz}
\usepackage{calc} 

\usetikzlibrary{arrows.meta, positioning}
\newcommand{\spin}{s}

\begin{document}


\title{Dynamical axion fields coupled with one-dimensional spinless fermions}

\author{Yuto Hosogi$^\text{1}$, Koichiro Furutani$^\text{1, 2}$, and Yuki Kawaguchi$^\text{1, 3}$}
\affiliation{
    $^\text{1}$Department of Applied Physics, Nagoya University, Nagoya 464-8603, Japan\\
	$^\text{2}$Institute for Advanced Research, Nagoya University, Nagoya 464-8601, Japan\\
    $^\text{3}$Research Center for Crystalline Materials Engineering, Nagoya University, Nagoya 464-8603, Japan
}
\date{\today}

\begin{abstract}
We investigate coupled dynamics of spinless fermions on a one-dimensional lattice and spins on the links. 
When the hopping integral and the on-site potential of the fermions depend on the direction of the link spins, the low-energy effective theory predicts that the link spins behave as a dynamical axion field in 1+1 dimensions.
The axion field $\theta$ is coupled to the electric field $E$ as $\theta E$, through which the link spins rotate in response to the applied electric field or the chemical potential gradient for charge-neutral fermions.
This is the inverse phenomenon of Thouless pumping in the Rice-Mele model.
After analyzing the dynamics by approximating the link spins with the classical ones and utilizing the axion Lagrangian, we show the full-quantum dynamics using the tensor network method.
Even though we do not explicitly introduce the axion Lagrangian in solving the fermion-spin coupled many-body dynamics, the full-quantum results agree well with those with the classical spin approximation, including the dynamics of the axion field and fermion transport.
In addition, we find that the quantum correlation between spins accelerates the dynamics of axion fields as the suppression of the expectation values of the link spins allows them to rotate easily.
We also propose a possible experimental setup for cold-atomic systems to implement the Hamiltonian in this study. 
\end{abstract}

\maketitle
\section{Introduction}

One of the state-of-the-art techniques in atomic gas experiments is controlling gauge fields in lattice systems~\cite{Dalibard_RMP2011, Goldman_RPP2014, Goldman_PRX2014, Goldman_nphys2016, Aidelsburger_CRP2018}, which is nowadays further developed to simulate lattice gauge theories~\cite{Lewenstein_2012text, Wiese_AdP2013, Zohar_RPP2016, Zache_qst2018, Surace_PRX2020, Aidelsburger_PTRSA2022, Zohar_PTRSA2022, Halimeh_nphys2025}. 
Historically, the lattice gauge theory has been primarily explored in elementary particle physics~\cite{Weinberg_1996, Rothe_2012, CarmenBanuls_RPP2020}. 
The key concept is to place interacting particles on lattice sites and attach gauge fields on the link, which defines a unitary transformation acting on the particle moving through the link. 
In atomic gas experiments, various configurations of gauge fields have been created by manipulating the hopping integral for atoms in an optical lattice using laser-assisted tunneling or shaking the lattice~\cite{Dalibard_RMP2011, Goldman_RPP2014, Goldman_PRX2014, Goldman_nphys2016, Aidelsburger_CRP2018}. 
Representative examples include the realization of topological phases in the Haldane and Hofstadter models~\cite{Miyake_PRL2013, Aidelsburger_PRL2013, Jotzu_nature2014, Aidelsburger_nphys2015}. 
We can also introduce atoms of a different species or in a different internal state to make the hopping of the original atoms dependent on the state of the added atoms, creating a dynamical gauge field~\cite{Lewenstein_2012text, Wiese_AdP2013, Zohar_RPP2016, Zache_qst2018, Surace_PRX2020, Aidelsburger_PTRSA2022, Zohar_PTRSA2022, Halimeh_nphys2025}. 
So far, the building blocks of the density-dependent gauge field~\cite{Clark_PRL2018, Gorg_NatPhys2019}, the $\mathbb{Z}_2$ gauge field~\cite{schweizer_nphys2019}, the four-body ring-exchange interactions~\cite{dai_nphys2017}, and the U(1) gauge field~\cite{Mil_Science2020} have been experimentally implemented after numerous theoretical works~\cite{Buchler_PRL2005, Osterloh_PRL2005, Cirac_PRL2010, Kapit_PRA2011, Zohar_PRL2011, Banerjee_PRL2012, Zohar_PRL2013, Tagliacozzo_NatComm2013, Bermudez_NJP2015, Kasper_NJP2017, Kuno_PRL2017, Barbiero_SciAdv2019, Halimeh_PRXQ2022}. Furthermore, the dynamical U(1) gauge field in a one-dimensional (1D) system has recently been realized, demonstrating Gauss's law in one dimension, thermalization dynamics in the gauge theory, and microscopic confinement dynamics by the topological $\theta$-term~\cite{Yang_nature2020, Zhou_Science2022, Zhang_NatPhys2025}. 
These systems serve as quantum simulators for high-energy physics. 
Axion physics is one of the examples that bridge topological phenomena in condensed matter and high-energy physics.

Originally, the axion field in 3+1 dimensions was proposed to solve the strong CP problem in quantum chromodynamics~\cite{PecceiQuinn_PRL1977, Weinberg_PRL1978, Wilczek_PRL1978}. 
The axion field $\theta$ couples with photons and introduces the so-called axion term, $\theta\bm E\cdot \bm B$, in the electromagnetic Lagrangian, where $\bm E$ and $\bm B$ are the electric and magnetic fields, respectively. 
The axion is now a candidate for dark matter, but it has yet to be directly observed~\cite{PRESKILL_PLB1983, ABBOTT_PLB1983, DINE_PLB1983}. 
On the other hand, we can assess the electromagnetic effects due to the axion term in topological materials~\cite{Qi_PRB2008, Li_NatPhys2010, Fradkin_2013, Nenno_NatRevPhys2020, Sekine_JAP2021}. 
In this context, the axion field is related to the band structure or the magnetic property of the material rather than an elementary particle. 
In 3D time-reversal invariant and inversion symmetric topological insulators, the axion field $\theta$ takes a quantized value $\pi$ (mod $2\pi$), leading to the quantized magnetoelectric effect~\cite{Mogi_NatMat2017, Mogi_SciAdv2017, Xiao_PRL2018} and the quantized Faraday and Kerr effects~\cite{Okada_NatCom2016, Wu_Science2016, Dziom_NatCom2017}. 
The value of $\theta$ can be arbitrary when the time-reversal and inversion symmetries are broken in axion insulators~\cite{Qi_PRB2008, Essin_PRL2009, Coh_PRB2011}. 
Furthermore, the axion field in the Weyl semimetals continuously changes in spacetime, giving rise to the anomalous Hall effect and the chiral magnetic effect in the bulk~\cite{Fukushima_PRD2008, Grushin_PRD2012, Zyuzin_PRB2012, Son_PRL2012, Wang_PRB2013, Vazifeh_PRL2013, Goswami_PRB2013, Zhou_CPL2013, Burkov_JPCM2015}. 
Although the axion fields in these materials are fixed or uniquely determined by the locations of the Weyl nodes, they can dynamically change in the presence of magnetic fluctuations~\cite{Li_NatPhys2010}. 
Accordingly, axion electrodynamics, i.e., the coupled dynamics of the axion field and the electromagnetic fields, has been theoretically investigated~\cite{Ooguri_PRL2012, Wang_PRB2016, Sekine_PRL2016, Sekine_PRB2016, Taguchi_PRB2018, Imaeda_JPSJ2019, Zhang_CPL2020, Xiao_PRB2021, Zhu_PRB2022, Xiao_PRB2023}. 
The axion field in 1+1 dimensions, which we discuss in this paper, is the low-dimensional version of the axion field in 3+1 dimensions. The corresponding axion Lagrangian is given by $\theta E$~\cite{Fradkin_2013}. 
The confinement dynamics of spinless fermions in 1+1 dimensions under a dynamical U(1) gauge with a tunable $\theta$-term has recently been investigated theoretically and experimentally~\cite{Halimeh_PRXQ2022, Zhang_NatPhys2025}. 

Dynamical axion fields in 1+1 dimensions are accessible in a dynamical Rice-Mele (RM) model~\cite{RiceMele_PRL1982, Asboth_2016text}. 
That is motivated by a previous work on the dynamical $\mathbb{Z}_2$ gauge field~\cite{Gonzalez_PRL2018}, where Ising spins are placed on every link between particle sites in a 1D optical lattice and the particle's hopping integral depends on the $z$ component of the spin. 
When hard-core bosons or fermions are loaded onto this lattice at half-filling, the antiferromagnetic order appears in the spin due to the Peierls transition. 
We focus on the fact that the antiferromagnetic spin order creates the Su–Schrieffer–Heeger (SSH) lattice for the particles~\cite{SSH_PRL1979}, which can become a dynamical RM model by adding an on-site potential dependent on the $x$ component of the link spin. In this case, the link spin forms a 1+1-dimensional axion field, allowing us to study the coupled dynamics of the axion and particle fields.

In this paper, we numerically investigate the many-body dynamics of fermions in a 1D optical lattice and spins on the links. 
The link spins introduce the $z$-component-dependent hopping integral and the $x$-component-dependent on-site potential for the fermions. 
Under the properly chosen parameters, the link spins antiferromagnetically order in the $x$-$z$ plane and create the RM model for the fermions. 
In the classical spin approximation, the direction of the antiferromagnetic spins in the $x$-$z$ plane corresponds to the axion field $\theta$, which dynamically evolves in time. 
Thus, when we apply a chemical potential gradient corresponding to the electric field $E$, the axion field evolves according to the axion Lagrangian $\theta E$, resulting in the rotation of the antiferromagnetic spins. 
One can interpret this dynamics as the inverse phenomenon of Thouless pumping~\cite{Thouless_PRB1983, Nakajima_NatPhys2016, Lohse_NatPhys2016}, where the periodic modulation of the parameters in the RM Hamiltonian induces a shift of fermions. 
In the present case, the flow of fermions under a chemical potential gradient causes the dynamics of RM parameters. 
The previous work~\cite{Sekine_PRB2016} discussed a similar phenomenon for topological magnetic insulators and predicted the electric-field-induced spin resonance as the inverse process of the dynamical chiral magnetic effect from the axion Lagrangian. 
Our study numerically demonstrates that such dynamics occur in the many-body system without explicitly introducing the axion Lagrangian. 
We prepare a stable staggered spin state and investigate the dynamics after suddenly applying a chemical potential gradient using the time-evolving block decimation (TEBD) method~\cite{Vidal_PRL2003, Vidal_PRL2004, Vidal_PRL2007, Banuls_PoS2019}. 
The numerical result agrees well with the classical spin dynamics, which obeys the Euler-Lagrange equation derived from the axion Lagrangian, supporting the validity of the axion Lagrangian. 

This paper is organized as follows. 
In Sec.~\ref{Secmodel}, we introduce the fermion-spin coupled system on a 1D lattice and explain how the system relates to the SSH and RM models. 
In Sec.~\ref{sec:reviewRM}, we briefly review the topological properties of the RM model and introduce the axion field in our system. 
In Sec.~\ref{sec:mean-field}, we derive the equation of motion of the classical spins from the axion Lagrangian, where we assume that the spins are approximated with the classical ones and the half-filled fermions fully occupy the lower band. 
By solving the derived equation of motion, we find that the applied chemical potential gradient, corresponding to the electric field for charged particles, induces the axion dynamics, resulting in the rotation of the classical spins. 
In Sec.~\ref{Sec:num}, we numerically confirm that the results in Sec.~\ref{sec:mean-field} hold without introducing the axion Lagrangian by solving the many-body dynamics using the TEBD method. 
In Sec.~\ref{Secexp}, we discuss a possible experimental setup to realize our model. 
Section~\ref{Secsummary} summarizes the results in this study. 
We set $\hbar=1$ throughout this paper. 

\section{Dynamical lattice in 1D}\label{Secmodel}
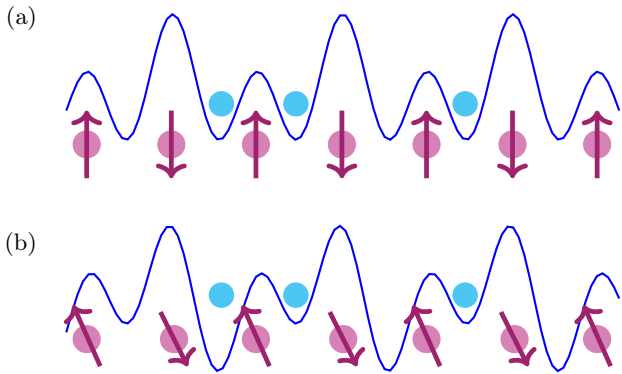
\begin{figure}[tbp]
    \centering
        \begin{minipage}[b]{1.0\columnwidth}
		\begin{tikzpicture}
            \node [below left] at (-2.8, 1.5) {(a)};
            \begin{scope}[shift={(0.3, 0)}, scale=0.9]
            
                \draw [blue, thick, domain=-pi:1.6*pi, samples=100] plot (\x, {+0.3*sin(2.5*\x r)-0.7*sin(5*\x r)-0.3*cos(2.5*\x r) + 0.3})node[right]{};

                \draw [fill, opacity = 0.7, cyan] (2.75, 0.1) circle [radius = 0.18];
                \draw [fill, opacity = 0.7, cyan] (0.25, 0.1) circle [radius = 0.18];
                \draw [fill, opacity = 0.7, cyan] (-0.85, 0.1) circle [radius = 0.18];

                \draw [fill, opacity = 1.0, Thistle] (4.7, -0.5) circle [radius = 0.2];
                \draw [<-, line width= 1.8pt, color = RedViolet](4.7, 0) -- (4.7, -1);
                %
                \draw [fill, opacity = 1.0, Thistle] (3.45, -0.5) circle [radius = 0.2];
                \draw [->, line width= 1.8pt, color = RedViolet](3.45, 0) -- (3.45, -1);
                \draw [fill, opacity = 1.0, Thistle] (2.18, -0.5) circle [radius = 0.2];
                \draw [<-, line width= 1.8pt, color = RedViolet](2.18, 0) -- (2.18, -1);
                \draw [fill, opacity = 1.0, Thistle] (0.93, -0.5) circle [radius = 0.2];
                \draw [->, line width= 1.8pt, color = RedViolet](0.93, 0) -- (0.93, -1);
                \draw [fill, opacity = 1.0, Thistle] (-0.34, -0.5) circle [radius = 0.2];
                \draw [<-, line width= 1.8pt, color = RedViolet](-0.34, 0) -- (-0.34, -1);
                \draw [fill, opacity = 1.0, Thistle] (-1.59, -0.5) circle [radius = 0.2];
                \draw [->, line width= 1.8pt, color = RedViolet](-1.59, 0) -- (-1.59, -1);
                \draw [fill, opacity = 1.0, Thistle] (-2.84, -0.5) circle [radius = 0.2];
                \draw [<-, line width= 1.8pt, color = RedViolet](-2.84, 0) -- (-2.84, -1);

        
            \end{scope}
            \node [below left] at (-2.8, -1.5) {(b)};
            \begin{scope}[shift={(0.3, -2.5)}, scale=0.9]
                \draw [blue, thick, domain=-pi:1.6*pi, samples=100] plot (\x, {-0.7*sin(5*\x r)+0.5*sin(2.5*\x r)})node[right]{};
                \draw [fill, opacity = 0.7, cyan] (2.75, 0.05) circle [radius = 0.18];
                \draw [fill, opacity = 0.7, cyan] (0.25, 0.05) circle [radius = 0.18];
                \draw [fill, opacity = 0.7, cyan] (-0.85, 0.05) circle [radius = 0.18];
    
                \draw [fill, opacity = 1.0, Thistle] (4.7, -0.6) circle [radius = 0.2];
                \draw [<-, line width= 1.8pt, color = RedViolet](4.5, -0.1) -- (4.9, -1);
                \draw [fill, opacity = 1.0, Thistle] (3.48, -0.6) circle [radius = 0.2];
                \draw [->, line width= 1.8pt, color = RedViolet](3.28, -0.2) -- (3.68, -1);
                \draw [fill, opacity = 1.0, Thistle] (2.18, -0.6) circle [radius = 0.2];
                \draw [<-, line width= 1.8pt, color = RedViolet](1.98, -0.1) -- (2.38, -1);
                \draw [fill, opacity = 1.0, Thistle] (0.95, -0.6) circle [radius = 0.2];
                \draw [->, line width= 1.8pt, color = RedViolet](0.75, -0.2) -- (1.15, -1);
                \draw [fill, opacity = 1.0, Thistle] (-0.34, -0.6) circle [radius = 0.2];
                \draw [<-, line width= 1.8pt, color = RedViolet](-0.54, -0.1) -- (-0.14, -1);
                \draw [fill, opacity = 1.0, Thistle] (-1.55, -0.6) circle [radius = 0.2];
                \draw [->, line width= 1.8pt, color = RedViolet](-1.75, -0.2) -- (-1.35, -1);
                \draw [fill, opacity = 1.0, Thistle] (-2.84, -0.6) circle [radius = 0.2];
                \draw [<-, line width= 1.8pt, color = RedViolet](-3.04, -0.1) -- (-2.64, -1);
            \end{scope}
        \end{tikzpicture}
    \end{minipage}
    \caption{
    Schematic of fermions (light blue filled circles) in a one-dimensional lattice coupled with spins (purple filled circles with arrows) on the links, described by Hamiltonian~\eqref{hamiltonian}. The $z$ component of spins determines the hopping amplitude of a fermion across the spin site (a), and the $x$ component of the spin modifies the on-site potential for a fermion at the site to the right (b).
    Blue curves depict the effective potential for the fermions.
    }
    \label{main_model}
\end{figure}

We start from a system of spinless fermions in a 1D dynamical lattice. The Hamiltonian is composed of three terms:
\begin{align}
    \hat{\mathcal{H}}' = \hat{\mathcal{H}}_{\mathbb{Z}_2}+\hat{\mathcal{H}}_{\text{F}}+\hat{\mathcal{H}}_{\Delta}.
    \label{hamiltonian_All}
\end{align}
The first term, $\hat{\mathcal{H}}_{\mathbb{Z}_2}$, describes the coupling between fermions on lattice sites and spins on links (Fig.~\ref{main_model}) and is given by
\begin{align}
    \hat{\mathcal{H}}_{\mathbb{Z}_2}
    =& -\alpha\sum_{j=1}^{L-1}\left(\hat{c}^{\dagger}_j\sigma^{z}_{j, j+1}\hat{c}_{j+1} + \text{h.c.}\right)
    +\beta\sum_{j=1}^{L-1}\sigma^{x}_{j,j+1},
    \label{hamiltonian_Z2}
\end{align}
where $\hat{c}^{\dagger}_j$ and $\hat{c}_j$ are the creation and annihilation operators of fermions at site $j$, respectively, 
$\sigma^{x,y,z}_{j, j+1}$ denotes the spin operator located on the link between sites $j$ and $j+1$, and $L$ is the number of lattice sites. The coefficients $\alpha$ and $\beta$ are real and assumed to be positive without loss of generality.
Here, because $\hat{\mathcal H}_{\mathbb{Z}_2}$ commutes with the generator of the $\mathbb{Z}_2$ local gauge transformation $\hat{G}_j \equiv\sigma_{j-1, j}^{x}(-1)^{\hat{n}_j}\sigma_{j, j+1}^{x}$ with $\hat{n}_j=\hat{c}_j^\dagger \hat{c}_j$, the spins on the links implement the $\mathbb{Z}_2$ Ising gauge fields acting on the fermions on the lattice sites. The confinement of fermions into bosonic dimers under the Hamiltonian~\eqref{hamiltonian_Z2} was predicted in Ref.~\cite{PhysRevLett.124.120503}.
The $\mathbb{Z}_2$ local gauge symmetry is preserved in the presence of the additional fermionic Hamiltonian
\begin{align}
    \hat{\mathcal{H}}_{\text{F}}
    = -t_h\sum_{j=1}^{L-1}\left(\hat{c}^{\dagger}_j\hat{c}_{j+1} + \text{h.c.}\right)
    -\mu\sum^{L}_{j=1}\hat{c}^{\dagger}_j\hat{c}_j,
\label{eq:Hamiltonian_fermion}
\end{align}
where $t_h(>0)$ is the hopping amplitude and $\mu$ is the chemical potential.
On the other hand, the Zeeman term
\begin{equation}
    \hat{\mathcal{H}}_{\Delta} =\frac{\Delta}{2}\sum_{j=1}^{L-1}\sigma_{j, j+1}^{z} 
    \label{gauss_Z2}
\end{equation}
corresponding to the longitudinal magnetic field $\Delta$ violates the local gauge symmetry.
In the following, we consider a nonzero $\Delta$.

With the fermions at half-filling and properly chosen $\Delta$ and $\beta$, the spins under the Hamiltonian~\eqref{hamiltonian_All} exhibit antiferromagnetic order along the $z$ direction due to the Peierls transition: Doubling the spatial periodicity accompanied by the antiferromagnetic order opens a band gap of fermions and reduces the total energy of the system~\cite{Gonzalez_PRL2018}. 
Under antiferromagnetic spins, 
the effective Hamiltonian for fermions reduces to the one for the Su–Schrieffer–Heeger (SSH) model~\cite{SSH_PRL1979, Asboth_2016text}:
\begin{equation}
\begin{split}
    \hat{\mathcal{H}}_{\text{SSH}}=
    &-v\sum_{j=1}^{L/2}\left(\hat{a}^{\dagger}_{j, \mathrm{A}}\hat{a}_{j, \mathrm{B}} + \text{h.c.}\right)\\
    &-w\sum_{j=1}^{L/2-1}\left(\hat{a}^{\dagger}_{j, \mathrm{B}}\hat{a}_{j+1, \mathrm{A}} + \text{h.c.}\right),
\end{split} 
 \label{hamiltonian_ssh}
\end{equation}
where we have introduced the sublattice index $(\mathrm{A},\mathrm{B})$ and redefined the fermion operators as $\hat{a}_{j,\mathrm{A}}=\hat{c}_{2j-1}$ and $\hat{a}_{j,\mathrm{B}}=\hat{c}_{2j}$. 
The hopping coefficients $v$ and $w$ vary depending on the spin state of the link. 
If spins are the eigenstates of $\sigma^z$ at each link with the alternative appearance of the eigenvalues $-1$ and $1$, they are given by $t_h\pm\alpha$.
The SSH model is a representative example of topological insulators~\cite{Asboth_2016text}: Hamiltonian~\eqref{hamiltonian_ssh} has topological edge states for $v<w$ (see also the next section). 
Note that because $v$ and $w$ in the present system are dynamical variables determined by the spin states, we can investigate the lattice-fermion coupled dynamics.
For example, for the case of the $\mathbb{Z}_2$ Bose-Hubbard model at half-filling,
an additional boson is predicted to be fractionalized into two, each confined in an accompanying magnetic domain wall~\cite{Gonzalez_PRL2020}.
\par

An interesting extension of the SSH model is the addition of a sublattice-dependent on-site potential, called the Rice-Mele (RM) model \cite{RiceMele_PRL1982, Asboth_2016text}, where periodic modulation of the parameters in the Hamiltonian induces topological charge pumping. 
To implement the dynamical RM model, we introduce an on-site potential for fermions dependent on the $x$ component of one of the neighboring spins: 
\begin{align}
	\hat{\mathcal{H}}
    = \hat{\mathcal{H}}' -\nu\sum_{j=2}^{L}\sigma^x_{j-1, j}\hat{c}^{\dagger}_{j}\hat{c}_{j} .
    \label{hamiltonian}
\end{align}
In the following, we consider spin-fermion coupled dynamics under Hamiltonian~\eqref{hamiltonian}. 
We fix the fermion fraction at half-filling to realize the sublattice spin structure. 
Before moving on to a detailed discussion on the dynamics, we briefly review the properties of the RM model in the next section. 

\section{Review of the Rice-Mele model and the axion field}
\label{sec:reviewRM}
\subsection{Thouless pumping in the Rice-Mele model}
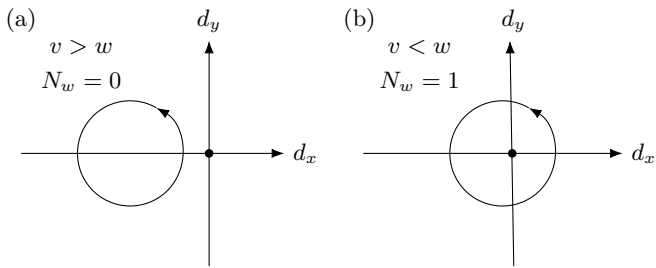
\begin{figure}[htbp]
    \centering
    \begin{tikzpicture}
            \draw[->, -{Latex}](-4, 0)--(-0.5, 0);
            \draw[->, -{Latex}](-1.5, -1.5)--(-1.5, 1.5);
            \draw [fill, black] (-1.5, 0) circle [radius = 0.05];
            \draw[->, -{Latex}](0.55, 0)--(4, 0);
            \draw[->, -{Latex}](2.55, -1.5)--(2.5, 1.5);
            \draw [fill, black] (2.53, 0) circle [radius = 0.05];
		\node [right] at (-0.5, 0) {$d_x$}; 
            \node [above] at (-1.5, 1.5) {$d_y$}; 
            \node [right] at (4, 0) {$d_x$}; 
            \node [above] at (2.55, 1.5) {$d_y$}; 
            \draw [->, -{Latex}] (2.75, 0.60621) arc [start angle=60, end angle=420, radius=0.7];
            \draw [->, -{Latex}] (-2.2, 0.60621) arc [start angle=60, end angle=420, radius=0.7];
             \node [above] at (-4, 1.5) {(a)}; 
             \node [above] at (0.5, 1.5) {(b)}; 
            
            \node [above] at (1.3, 1.2) {$v<w$}; 
            \node [below] at (1.3, 1.2) {$N_w=1$}; 
            \node [above] at (-3.2, 1.2) {$v>w$}; 
            \node [below] at (-3.2, 1.2) {$N_w=0$}; 
    \end{tikzpicture}
    \caption{
    Trajectories of vector ${\bm d}(k)$ defined in Eq.~\eqref{eq:H_SSH} for $v>w$ (a) and $v<w$ (b), which have the winding number $N_w=0$ and $1$, respectively,  according to Eq.~\eqref{eq:winding}.
    }
\label{fig:SSH_charge}
\end{figure}

We start with a recap of the topological properties of the SSH model.
By assuming the periodic boundary condition, the SSH Hamiltonian~\eqref{hamiltonian_ssh} is rewritten in the momentum space as
\begin{align}
\hat{\mathcal H}_\textrm{SSH}&=\sum_k \begin{pmatrix}\hat{a}^\dagger_{k,\mathrm{A}}& \hat{a}^\dagger_{k,\mathrm{B}}\end{pmatrix}
H_\textrm{SSH}(k)
\begin{pmatrix}\hat{a}_{k,\mathrm{A}}\\ \hat{a}_{k,\mathrm{B}}\end{pmatrix},\\
H_\textrm{SSH}(k)&=\begin{pmatrix}
0 & -v-we^{-ik}\\
-v-we^{ik} & 0
\end{pmatrix}= \bm d(k)\cdot\bm\tau,
\label{eq:H_SSH}
\end{align}
where $\bm\tau=(\tau^x,\tau^y,\tau^z)$ is a vector of Pauli matrices in the sublattice space, $\bm{d}(k)=(-v-w\cos k, -w\sin k, 0)$, and $\hat{a}_{k,\mathrm{A/B}}=\sum_{j=1}^{L/2} \hat{a}_{j,\mathrm{A/B}} \,e^{-ikj}$. 
Under the chiral symmetry in the SSH model, i.e., $\{ H_\textrm{SSH}(k),\tau^z\}=0$ for $\forall k\in [-\pi,\pi]$, the topological phase is classified by the winding number
\begin{align}
N_w=\frac{i}{2\pi}\int_{-\pi}^\pi \dd k q^*\dv{k}q,
\label{eq:winding}
\end{align}
with $q=(d_x+id_y)/\sqrt{d_x^2+d_y^2}$. 
The winding number corresponds to the times the vector $\bm{d}(k)$ circumnavigates around the origin in the $d_x$-$d_y$ plane during $k$ changes from $-\pi$ to $\pi$.
Hence, for $\bm d(k)$ defined in Eq.~\eqref{eq:H_SSH}, we obtain $N_w=1$ ($N_w=0$) for $v<w$ ($v>w$)
(see Fig.~\ref{fig:SSH_charge}). 
Consequently, the topological edge state arises (does not arise) for $v<w$ ($v>w$). 

In the RM model, the Hamiltonian has an additional sublattice-dependent on-site potential,
\begin{align}
    \hat{\mathcal{H}}_{\text{RM}}=&
    \hat{\mathcal H}_\textrm{SSH}
    +u\sum_{j=1}^{L/2}\left(\hat{a}^{\dagger}_{j, \mathrm{A}}\hat{a}_{j, \mathrm{A}} - \hat{a}^{\dagger}_{j, \mathrm{B}}\hat{a}_{j, \mathrm{B}}\right),
\end{align}
which introduces the $z$ component of $\bm d(k)$:
\begin{align}
    \hat{\mathcal{H}}_{\text{RM}}
    =&\sum_k \begin{pmatrix}\hat{a}^\dagger_{k,\mathrm{A}}& \hat{a}^\dagger_{k,\mathrm{B}}\end{pmatrix}H_\textrm{RM}(k)
    \begin{pmatrix}\hat{a}_{k,\mathrm{A}}\\ \hat{a}_{k,\mathrm{B}}\end{pmatrix},\\
    H_\textrm{RM}(k)=&-(v+w\cos k )\tau^x - w\sin k \tau^y+u\tau^z.
\label{hamiltonian_ricemele}
\end{align}
Due to the finite $\tau^z$ term, $H_\textrm{RM}(k)$ no longer anticommutes with $\tau^z$ and breaks the chiral symmetry. 
The broken chiral symmetry allows us to continuously deform $H_\textrm{SSH}(k)$ from the topological to the trivial phases without closing the energy gap. 
For instance, suppose that the parameters in the Hamiltonian evolve in time as
\begin{equation}
u=\sin{(\Omega t)}, \quad 
v=1-\cos{(\Omega t)}, \quad 
w=1,
\end{equation}
where we choose the frequency $\Omega$ small enough such that the system adiabatically follows the change in the Hamiltonian.
When $t$ is an integer multiple of $\pi/\Omega$, $H_\textrm{RM}(k)$ reduces to $H_\textrm{SSH}(k)$, at which the winding number~\eqref{eq:winding} is well defined and alternatively takes $1$ and $0$.
Consequently, the topological edge state periodically appears, resulting in charge pumping.
In the case of solid-state electrons constituted from many bands, this oscillation corresponds to the shift of electric polarization inside the sublattice~\cite{Asboth_2016text}.
On the other hand, in the pristine RM model, one periodic cycle shifts the fermions' distribution by one sublattice, and multiple cycles move the center of mass in one direction, as demonstrated in Refs.~\cite{Nakajima_NatPhys2016, Lohse_NatPhys2016}.

\subsection{Axion field}
\label{sec:axion_field}
When the parameters $u, v$, and $w$ are not externally controllable parameters but dynamical parameters determined self-consistently, the change of the fermionic state in response to external electromagnetic fields can induce the dynamics of the parameter field, called the axion field. 
In this section, we introduce the axion field and the corresponding action within the Lagrangian formalism. 

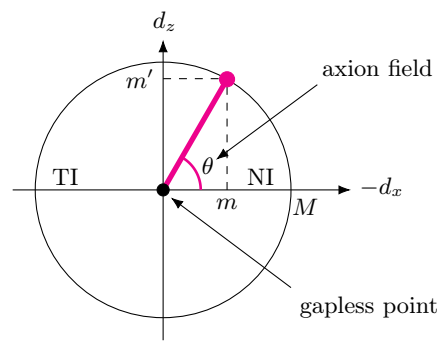
\begin{figure}[htbp]
    \centering
    \begin{tikzpicture}
		\draw (0, 0) circle [radius = 1.7];
		\node [above right] at (2.5, -0.2) {$-d_x$}; 
		\draw [->, -{Latex}](-2, 0) -- (2.5, 0);
		\node [above left] at (0.3, 2) {$d_z$}; 
		\draw [dashed](0, 1.48) -- (0.85, 1.48);
		\draw [dashed](0.85, 0) -- (0.85, 1.48);
		\node [below] at (0.85, 0) {$m$}; 
		\node [left] at (0, 1.48) {$m'$}; 
		\node [below right] at (1.6, 0) {$M$}; 
		\draw [->, -{Latex}](0, -2) -- (0, 2);
		\draw [domain = 0:60, line width=1pt, smooth, variable=\t, magenta] plot({0.5 * cos(\t)}, {0.5 * sin(\t)});
		\node [above right] at (0.4, 0.1) {$\theta$}; 
		\node [above right] at (2.0, 1.4) {axion field}; 
		\draw [<-, {Latex}-] (0.7, 0.4) -- (2.05, 1.4);
		\draw [color = magenta, line width = 2pt](0, 0) -- (0.85, 1.472);
		\draw [fill, magenta] (0.85, 1.472) circle [radius = 0.1];
		\draw [fill, black] (0, 0) circle [radius = 0.08];
		\node [below right] at (1.7, -1.3) {gapless point}; 
		\draw [->, {Latex}-] (0.1, -0.1) -- (1.7, -1.3);
		\node [above] at (1.3, -0.05) {NI}; 
		\node [above] at (-1.3, -0.05) {TI}; 
    \end{tikzpicture}
    \caption{
    Schematic for the relation between the mass terms $-d_x=m$ and $d_z=m'$ in the Hamiltonian \eqref{eq:effective_Hk} and the axion field $\theta$.
    The topological insulator (TI) and normal insulator (NI) phases of the SSH model lie on the negative and positive horizontal axes, respectively.
    In the RM model, the system can continuously change from the TI to the NI phases without closing a gap by introducing the vertical axis.
    Such a motion induces the charge pumping in the RM model.
    When the chiral gauge transformation rotates the system in the ($-d_x$)-$d_z$ space, an additional Lagrangian for the axion field $\theta$ arises as in Eqs.~\eqref{eq:Lag_tot} and \eqref{eq:Lag_axion}.
    }
\label{Axion_rotation}
\end{figure}

Based on the smallest energy split at $k=\pi$ in $H_\textrm{RM}(k)$, we expand $H_\textrm{RM}(\pi+k)$ with respect to $k$ and construct the effective theory for a small $k$. 
The low-energy effective Hamiltonian reduces to
\begin{align}
H(k)=-m\tau^x+k\tau^y+m'\tau^z-\mu,
\label{eq:effective_Hk}
\end{align}
up to the first order in $k$ where we have introduced the chemical potential $\mu$, rescaled the length scale as $wk\to k$, and defined $m=v-w$, and $m'=u$. 
Note that the next-order term to Eq.~\eqref{eq:effective_Hk} is $-k^2/(2w)\tau^x$, which is necessary to define the winding number at $m'=0$:
When $k$ changes from $-\infty$ to $\infty$, $\arg(q)$ in Eq.~\eqref{eq:winding} changes by $2\pi$ ($0$) for $m<0$ ($m>0$), leading to $N_w=1$ ($N_w=0$).
On the other hand, when we ignore the $k^2$ term, Eq.~\eqref{eq:effective_Hk} is continuously transformed to the one for the trivial phase of the SSH model, i.e., the normal insulator, 
\begin{align}
H_\textrm{NI}(k)=-M\tau^x+k\tau^y-\mu,
\label{eq:effective_Hk_SSH}
\end{align}
with  $M=\sqrt{m^2+(m')^2}$
under the rotation of the sublattice basis around the $d_{y}$ axis by the angle
\begin{align}
\theta=\arg(m+im'),
\label{eq:def_axion}
\end{align}
which is equivalent to the rotation in the $-d_x$-$d_z$ space from $(-d_x,d_z)=(m,m')$ to $(M,0)$ as shown in Fig.~\ref{Axion_rotation}.
However, this rotation introduces an additional Lagrangian concerning $\theta$, the axion field.

To see this, we move on to the Lagrangian description.
With the inverse Fourier transformation $\hat{\Psi}_{\rm X}=(2/L)\sum_k e^{ikx}\hat{a}_{k, \mathrm{X}}$ ($\mathrm{X} = \mathrm{A, B}$), the Euclidean action of the fermions reads
\begin{align}
    S^{\text{E}}=& i\iint \dd t\dd x \mathcal{L}^{\text{E}}\nonumber \\
    =
    &\iint \dd\tau \dd x \begin{pmatrix} \hat{\Psi}_{\rm A}^\dagger & \hat{\Psi}_{\rm B}^\dagger\end{pmatrix}
    \left[\partial_\tau+H(-i\partial_x)\right]
    \begin{pmatrix} \hat{\Psi}_{\rm A} \\ \hat{\Psi}_{\rm B}\end{pmatrix},
    \label{action0}
\end{align}
where $\mathcal{L}^{\text{E}}$ is the Euclidean Lagrangian density, $\tau=it$ is the imaginary time, and $H(-i\partial_x)$ is the coordinate-space representation of Eq.~\eqref{eq:effective_Hk}. 

When we use Eq.~\eqref{eq:effective_Hk_SSH} instead of Eq.~\eqref{eq:effective_Hk}, the action is written in the same form as the Weyl action in 1+1 dimensions:
\begin{equation}
    S_{\rm NI}^{\text{E}} = \iint \dd[2]{x} \bar{\psi}\left[
        \gamma^{\varsigma}(\partial_{\varsigma} + iA_{\varsigma}) + M
    \right]\psi,
    \label{dirac_S} 
\end{equation} 
where $\psi \equiv (\hat{\Psi}_{\rm A}, \hat{\Psi}_{\rm B})^{\rm T}$ and $\bar{\psi}=\psi^{\dagger}\gamma^1$ are the spinor field and its adjoint, respectively, $\dd[2]{x}=\dd x_1\dd x_2$ with $x_1=\tau, x_2=x$, and repeated indices
imply the summation over them for 1 and 2 following the Einstein notation.
The corresponding Gamma matrices are defined by
$\gamma^1=-\tau^x, \gamma^2=-\tau^z$, and $\gamma^5=-i\gamma^1\gamma^2=-\tau^y$, and the electromagnetic potential is determined by the chemical potential as $A_1=i\mu$ and $A_2=0$. 

The action corresponding to Eq.~\eqref{eq:effective_Hk} is obtained by rotating the sublattice basis in Eq.~\eqref{dirac_S}, i.e., by replacing $\psi$ and $\bar{\psi}$ in Eq.~\eqref{dirac_S} with $\psi'$ and $\bar{\psi'}$ and substituting
\begin{align}
\psi'= e^{-i\theta\gamma^5/2}\psi,\ \ \ 
    \bar{\psi}'= \bar{\psi}e^{-i\theta\gamma^5/2}.
    \label{Chiral_Gauge}
\end{align}
This is the chiral gauge transformation for the spinor fields.
The invariance of the partition function under this transformation yields
\begin{align}
  \mathcal{Z}&
    =\int\mathcal{D}\psi\mathcal{D}\bar{\psi}e^{-S^{\text{E}}_\textrm{NI}[\psi, \bar{\psi}]}
    =\int \mathcal{D}\psi'\mathcal{D}\bar{\psi}'e^{-S^{\text{E}}_{\text{NI}}[\psi', \bar{\psi}']}\nonumber\\   
    &=\int J\mathcal{D}\psi \mathcal{D}\bar{\psi} e^{-S^{\text{E}}_{\text{NI}}[e^{-i\theta\gamma^5/2}\psi, \bar{\psi}e^{-i\theta\gamma^5/2}]}\nonumber\\
     &=\int \mathcal{D}\psi \mathcal{D}\bar{\psi} e^{-S^{\text{E}}[\psi, \bar{\psi}]+\ln J},
\end{align}
i.e.,
\begin{equation}
   S^{\text{E}}[\psi, \bar{\psi}]=S^{\text{E}}_\textrm{NI}[\psi, \bar{\psi}]+\ln J,
\end{equation}
with $J$ being the Jacobian of the transformation.
The Jacobian is practically obtained by applying multiple infinitesimal rotations and given by~\cite{Adler_PhysRev1969, Fujikawa_PRL1979, Fujikawa_PRD1980}
\begin{equation}
\begin{aligned}[b]
J&=\exp\left[-\int \dd[2]x \frac{i\theta}{4\pi}\epsilon_{\varsigma \upsilon}F_{\varsigma \upsilon}\right] \\
&=\exp\left[-\int \dd[2]x \frac{\theta}{2\pi} \pdv{\mu}{x}\right],
\end{aligned}
\end{equation}
where $\epsilon_{\varsigma \upsilon}$ is the antisymmetric tensor
and $F_{\varsigma \upsilon}=\partial_\varsigma A_\upsilon-\partial_\upsilon A_\varsigma$ is the Euclidean electromagnetic tensor. 
With the Euclidean Lagrangian densities ${\mathcal L}^\textrm{E}_\textrm{NI}$ and ${\mathcal L}^\textrm{E}_\textrm{Axion}$ corresponding to $S^\textrm{E}_\textrm{NI}[\psi,\bar{\psi}]$ and $\ln J$, respectively, we can divide the Lagrangian density of the system into the one for the fermions in the normal insulator and the one for $\theta$ as
\begin{align}
{\mathcal L}={\mathcal L}_\textrm{NI} + {\mathcal L}_\textrm{Axion},
\label{eq:Lag_tot}
\end{align}
where $\mathcal{L}=-\mathcal{L}^{\text{E}}$ is the Lagrangian density in Minkowski spacetime. 
Consequently, we obtain the axion Lagrangian density
\begin{align}
{\mathcal L}_\textrm{Axion}
=-\mathcal{L}^{\text{E}}_\textrm{Axion}
=\frac{\theta}{2\pi} \pdv{\mu}{x}.
\label{eq:Lag_axion}
\end{align}
Here, $\theta$ is the axion field in 1+1 dimensions.

In our model \eqref{hamiltonian}, the spins on the links constitute the axion field. 
Equation~\eqref{eq:Lag_tot} implies that, in describing the spin dynamics,
we can neglect the motional degrees of freedom of fermions 
by considering the coupling between the axion field and the chemical potential gradient via the axion Lagrangian. 
In this description, fermions are regarded as being in the normal insulating state. 
In the following, we investigate the spin dynamics induced by the chemical potential gradient. In Sec.~\ref{sec:mean-field}, we solve the equation of motion of spins derived from the axion Lagrangian under the classical spin approximation. 
In Sec.~\ref{Sec:num}, we numerically confirm that the result holds for quantum spins by directly solving the many-body dynamics following Hamiltonian~\eqref{hamiltonian}.

\section{Classical spin approximation}\label{sec:mean-field}
This section is devoted to the analysis of the spin dynamics within the classical spin approximation. We first show that the staggered spin state arises in the ground state for a certain parameter set. We then derive the equation of motion for the staggered spins from the Lagrangian including the axion term and investigate the spin dynamics in response to the applied chemical potential gradient.

\subsection{Staggered Spin Configuration}
\label{sec:staggered_spin}
We first search for the set of parameters that stabilizes a staggered spin configuration in both the $x$ and $z$ components as the ground state.
We suppose a two-sublattice spin structure and assume that the directions of spins on sublattices $\rm A$ and $\rm B$ are parametrized as 
\begin{subequations}
\label{eq:staggered_spin_ansatz}
\begin{align}
    (\spin_{\rm A}^x, \spin_{\rm A}^y, \spin_{\rm A}^z)
    &=(\sin\vartheta\cos\varphi, \sin\vartheta\sin\varphi, \cos\vartheta),\\
    &\equiv(s^x,s^y,s^z),\nonumber\\
    (\spin_{\rm B}^x, \spin_{\rm B}^y, \spin_{\rm B}^z)&=(-\sin\vartheta\cos\varphi, \sin\vartheta\sin\varphi, -\cos\vartheta),\\
    &=(-s^x,s^y,-s^z).\nonumber
    \end{align}
\end{subequations}
Without loss of generality, we can restrict $0\le \vartheta \le \pi/2$.
The staggered spin state in both the $x$ and $z$ components arises when the system has an energy minimum at $\vartheta\neq 0, \pi/2$. Energy comparisons with the other spin configurations are provided in Appendix~\ref{energy_minimums}. 

Under the appearance of the two-site sublattice spin structures, the Hamiltonian~\eqref{hamiltonian} is rewritten in momentum space as
\begin{widetext}
\begin{align}
     \hat{\mathcal{H}}=&\sum_k(\hat{a}_{k, \mathrm{A}}^{\dagger}\, \hat{a}_{k, \mathrm{B}}^{\dagger})H_k\begin{pmatrix}
        \hat{a}_{k, \mathrm{A}}\\
        \hat{a}_{k, \mathrm{B}}\\
    \end{pmatrix}+\frac{L}{2}\left[\beta(\spin^x_\mathrm{A}+\spin^x_\mathrm{B})
    +\frac{\Delta}{2}(\spin^z_\mathrm{A}+\spin^z_\mathrm{B})\right],
    \label{eq:Hall_in_k-space}
\end{align} 
where $H_k$ is the $2\times 2$ matrix defined by
\begin{align}
    H_k=
    \begin{pmatrix}
        -\mu-\nu\spin^x_{\rm A} & -(t_h+\alpha\spin^z_{\rm B}) -(t_h+\alpha\spin^z_{\rm A})e^{-ik}\\
        -(t_h+\alpha\spin^{z}_{\rm B})-(t_h+\alpha\spin^z_{\rm A})e^{ik} & -\mu-\nu\spin^x_{\rm B} \\
    \end{pmatrix}.
    \label{Hamiltonian_2_2}
\end{align}
The fermion band dispersion is obtained by diagonalizing $H_k$ as
\begin{align}
\epsilon_{\pm}(k)=-\mu-\nu\frac{\spin_\mathrm{A}^x+\spin_\mathrm{B}^x}{2}\pm \sqrt{\nu^2\left(\frac{\spin_\mathrm{A}^x-\spin_\mathrm{B}^x}{2}\right)^2+(t_h+\alpha\spin_\mathrm{A}^z)^2+(t_h+\alpha\spin_\mathrm{B}^z)^2+2(t_h+\alpha\spin_\mathrm{A}^z)(t_h+\alpha\spin_\mathrm{B}^z)\cos k}.
\label{eq:fermion_dispersion}
\end{align}
The fermion energy at half-filling is obtained by integrating the lower band in the first Brillouin zone.
The resulting energy per sublattice under the staggered spin configuration reads
    \begin{equation}
	\begin{aligned}[b]
        \frac{2E_\textrm{ss}(\vartheta, \varphi)}{L}
        &=-\mu-\int_{-\pi}^{\pi}\dfrac{\dd{k}}{2\pi}\sqrt{
            2t_h^2(1+\cos k)+2\alpha^2\cos^2\vartheta(1-\cos k)+\nu^2\sin^2\vartheta\cos^2\varphi
        }\\
        &=-\mu-\frac{4t_h}{\pi}\sqrt{1+\frac{\nu^2}{4t_h^2}\sin^2\vartheta\cos^2\varphi}\,{\mathfrak E}\left(\frac{1-\frac{\alpha^2}{t_h^2}\cos^2\vartheta}{1+\frac{\nu^2}{4t_h^2}\sin^2\vartheta\cos^2\varphi}\right),\label{Energy}
    \end{aligned}
    \end{equation}
\end{widetext}
where we have used Eq.~\eqref{eq:staggered_spin_ansatz}, and ${\mathfrak E}(\kappa)=\int_0^{\pi/2}\sqrt{1-\kappa\sin^2x}\,dx$ is the complete elliptic integral of the second kind.

Because ${\mathfrak E}(\kappa)$ is a monotonically decreasing function, $E_\textrm{ss}(\vartheta,\varphi)$ has a minimum at $\varphi=0$ as a function of $\varphi$. 
On the other hand, as a function of $\vartheta$, $E_\textrm{ss}(\vartheta,0)$ has a minimum at $\vartheta=0$ for a small $\nu/t_h$.
Increasing $\nu/t_h$ intensifies $\nu^2/(4t_h^2)\sin^2\vartheta$, and the optimal $\vartheta$ deviates from $0$.
At large enough $\nu/t_h$, $E_\textrm{ss}(\vartheta,0)$ takes its minimum at $\vartheta=\pi/2$.
In between, there is a window of $\nu$ such that $E_\textrm{ss}(\vartheta,0)$ takes its minimum at $\vartheta\neq 0,\pi/2$.

\begin{figure}[t]
	\centering
	\includegraphics[width=\columnwidth]{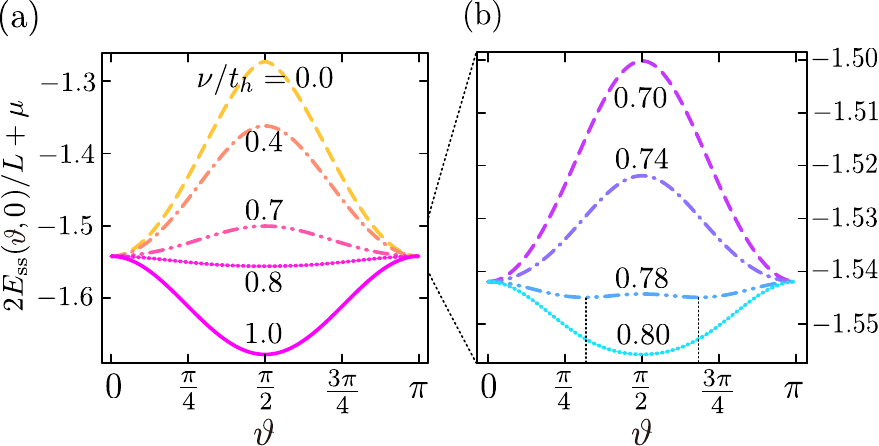}
	\caption{
		(a) Energy $E_\textrm{ss}(\vartheta,0)$ of the staggard spin configuration in the $x$ and $z$ directions as a function of $\vartheta$
	   for various values of $\nu/t_h$ at $\alpha/t_h=0.5$.
       (b) Magnified view of (a) for $0.7\le \nu/t_h\le 0.8$.
        At $\nu/t_h = 0.78$, $E_\textrm{ss}(\vartheta,0)$ has a minimum at $\vartheta \simeq \pi/3$. 
	}\label{Energy_ground}
\end{figure}

We numerically verify that such $\nu$ exists. 
Figure~\ref{Energy_ground} shows the $\vartheta$ dependence of $2E_\textrm{ss}(\vartheta,0)/L+\mu$ for several values of $\nu/t_h$ and fixed $\alpha/t_h=0.5$ with numerical evaluation of ${\mathfrak E}(\kappa)$.
As $\nu$ increases, the minimum point of $2E_\textrm{ss}(\vartheta,0)/L+\mu$ changes from $\vartheta=0$ to $\pi/2$, and in between we obtain an optimal $\vartheta\neq 0,\pi/2$, e.g., $\vartheta\simeq \pi/3$ at $\nu/t_h=0.78$.

For $E_\textrm{ss}(\vartheta,\varphi)$ being independent of $\Delta$ and $\beta$, the energy argument above in Fig.~\ref{Energy_ground} holds regardless of their choices. 
They are determined such that the anzats~\eqref{eq:staggered_spin_ansatz} has a lower energy than the other possible spin configurations, which yields $\beta=\nu/2$ and $\Delta\simeq 4\alpha/\pi$. See Appendix~\ref{energy_minimums} for details.

\subsection{Axion field in the dynamical RM model}

Following the discussion in Sec.~\ref{sec:axion_field}, we define the axion field for the Hamiltonian~\eqref{hamiltonian}.
We expand $H_k$ in Eq.~\eqref{Hamiltonian_2_2} around $k=\pi$ and rewrite $H_{\pi +k}$ in the form of Eq.~\eqref{eq:effective_Hk}.
Then, $m$ and $m'$ are written in terms of the spin directions as
\begin{align}
m&=-\alpha(\spin_{\rm A}^z-\spin_{\rm B}^z)=-2\alpha\cos\vartheta, \\
m'&=-\frac{\nu}{2}(\spin_{\rm A}^x-\spin_{\rm B}^x)=-\nu\sin\vartheta\cos\varphi,
\end{align}
which defines the axion field of Eq.~\eqref{eq:def_axion} as
\begin{align}
\theta=\arg(m+im')=\arg(2\alpha s^z+i\nu s^x)+\pi.
\label{eq:def_axion_spin}
\end{align}
Equation \eqref{eq:Lag_axion} is the corresponding Lagrangian density. 

\subsection{Spin dynamics under a chemical potential gradient}
We derive the equation of motion of spins from the Lagrangian.
We assume that the spins hold the two-site sublattice structure given by Eq.~\eqref{eq:staggered_spin_ansatz} during the dynamics.
The Lagrangian density per unit cell reads
\begin{align}
    \mathcal{L}= 2{\dv{\varphi}{t}} \cos\vartheta - \frac{2E_\textrm{ss}(\vartheta, \varphi)}{L}+ 
    \mathcal{L}_{\text{Axion}},
\label{Lagrangean_all}
\end{align}
Here, the first term is the sum of the Berry phase terms of two sublattices, which remains because of the ferromagnetic order in the $y$ direction~\cite{Fradkin_2013}.
The second term, given in Eq.~\eqref{Energy}, is the energy density of fermions filling the lower band, which depends on the spin direction and thus is necessary even though we neglect the motional degrees of fermions.  
In the axion Lagrangian density $\mathcal{L}_{\rm Axion}$, the chemical potential gradient $\partial \mu/\partial x$ corresponds to the chemical potential difference between the neighboring unit cells $\Delta\mu\equiv\mu_{j+2}-\mu_j$ where $\mu_j$ is the chemical potential for the fermion at site $j$. 

The Lagrangian \eqref{Lagrangean_all} yields the Euler-Lagrange equation for the staggered spins, which is summarized as
\begin{align}
	\dv{\bm{\spin}}{t}=&\begin{pmatrix}
		\alpha^2 E_2(\bm{\spin})\spin^y\spin^z\\
		\dfrac{\nu^2}{2}E_1(\bm{\spin})\spin^x\spin^z-\alpha^2 E_2(\bm{\spin})\spin^x\spin^z\\
		-\dfrac{\nu^2}{2}E_1(\bm{\spin})\spin^x\spin^y
	\end{pmatrix}\notag\\
	&+\Gamma(\bm{\spin})\begin{pmatrix}
		-\spin^x\spin^y \\ (\spin^x)^2 + (\spin^z)^2 \\ -\spin^y\spin^z
	\end{pmatrix},\label{spin_time_evo}
\end{align}
where $\bm \spin=(s^x,s^y,s^z)=(\sin\vartheta\cos\varphi, \sin\vartheta\sin\varphi, \cos\vartheta)$ denotes the spin direction on the sublattice $\rm A$,
and $E_1(\bm{\spin}), E_2(\bm{\spin})$, and $\Gamma(\bm{\spin})$ are, respectively, defined by
\begin{align}
	E_1&=
        \int_{-\pi}^{\pi}\frac{\dd{k}/2\pi}{\sqrt{
		2t_h^2(1+\cos k )+2(\alpha\spin^z)^2(1-\cos k)+(\nu\spin^x)^2
	}} ,\\
	E_2&=
            \int_{-\pi}^{\pi}\frac{(1-\cos k)\dd{k}/2\pi}{\sqrt{
		2t_h^2(1+\cos k )+2(\alpha\spin^z)^2(1-\cos k)+(\nu\spin^x)^2
	}} ,
\end{align}\vspace{-5mm}
\begin{align}
    	\Gamma=\frac{\Delta\mu}{2\pi}
	\frac{\alpha \nu }{(\nu\spin^x)^2 + (2\alpha\spin^z)^2}.
    \label{eq:Gamma}
\end{align}
Here, the first term in the right-hand side of Eq.~\eqref{spin_time_evo} represents the dynamics originating from the spin-dependent dispersion of the fermion band, while the second term is derived from the axion Lagrangian. 

\begin{figure*}[t]
 \begin{center}
  \includegraphics[width=0.95\linewidth]{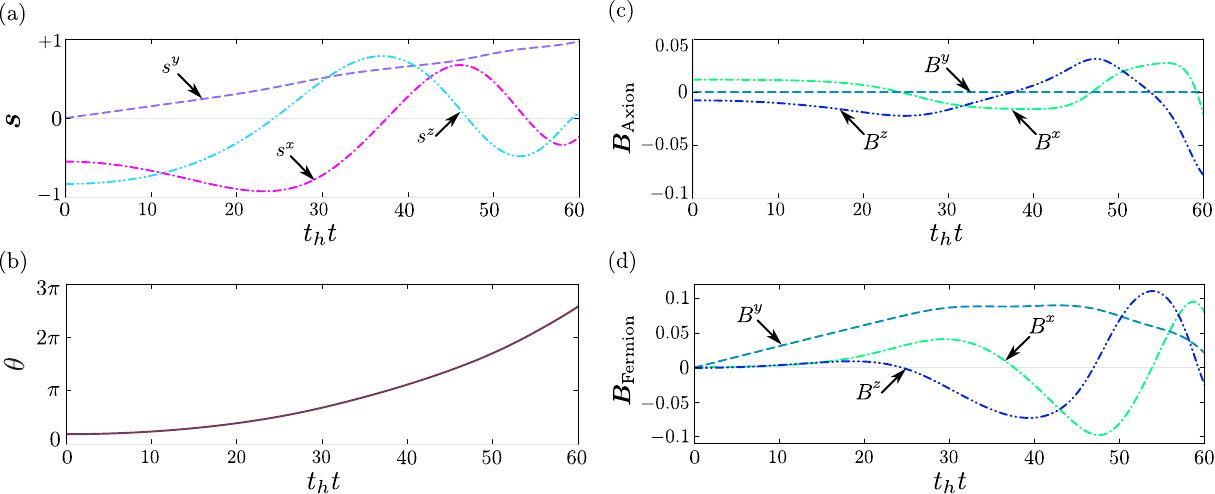}
  \caption{(a) Spin dynamics within the classical spin approximation with the two-sublattice staggered spin structure [Eq.~\eqref{eq:staggered_spin_ansatz}] obtained by numerically solving Eq.~\eqref{spin_time_evo} for $\alpha/t_h=0.5, \nu/t_h=0.775, \Delta\mu/t_h=0.2$ starting from the stationary state obtained at $\Delta\mu=0$. 
  (b) Time evolution of the axion field calculated from the spin component shown in (a) via Eq.~\eqref{eq:def_axion_spin}.
  (c)(d) Time evolutions of the effective magnetic field due to the axion field, $\bm B_\textrm{Axion}$ (c), and the one from the coupling with the fermions, $\bm B_\textrm{Fermion}$ (d). $\bm B_\textrm{Fermion}$ vanishes at $t=0$, whereas $\bm B_\textrm{Axion}$ at $t=0$ lies in the $x$-$z$ plane and triggers the spin dynamics as plotted in (a). 
}
  \label{fig_analytical_calculation}
 \end{center}
\end{figure*}

We numerically solve Eq.~\eqref{spin_time_evo} using the fourth-order Runge-Kutta method under the initial state as the stationary solution of Eq.~\eqref{spin_time_evo} at $\Delta \mu =0$,
which is given by the optimal $\vartheta$ and $\varphi=0$ discussed in Sec.~\ref{sec:staggered_spin}.
We then suddenly turn on $\Delta \mu$ at $t=0$ and see the subsequent dynamics.
Figure~\ref{fig_analytical_calculation}(a) shows the dynamics of $\bm\spin$ for $\alpha/t_h=0.5$, $\nu/t_h=0.775$, and $\Delta\mu/t_h=0.2$, with the initial state $\bm\spin=(-0.55, 0, -0.83)$, where we have chosen the parameters so that the initial $\theta$ coincides with the one in the numerical simulation in the next section. One can see that $\bm\spin$ rotates about the $y$ axis with a slowly increasing $y$ component. 
This rotation about the $y$ axis is the inverse phenomenon of Thouless pumping:
The particle flow caused by the chemical potential gradient induces rotation in the parameter space of the mass terms (see Fig.~\ref{Axion_rotation}).
We show the time evolution of the axion field $\theta$ in Fig.~\ref{fig_analytical_calculation}(b), which shows the monotonically increasing $\theta$ that reduces the axion energy $-\mathcal{L}_\textrm{Axion}$.

To see the role of the axion term in the dynamics, we introduce the effective magnetic field $\bm{B}_{\text{eff}}=\bm{\spin}\times {\mathrm d}{\bm{\spin}}/\mathrm{d}{t}$ and divide it into $\bm{B}_{\rm Fermion}$ and $\bm{B}_{\rm Axion}$, corresponding to the first and second terms of Eq.~\eqref{spin_time_evo}, respectively.
Here, $\bm B_\textrm{Axion}$ is always perpendicular to the $y$ axis as
\begin{align}
\bm B_\textrm{Axion}=\Gamma(\bm\spin)\begin{pmatrix} -s^z \\ 0 \\ s^x\end{pmatrix}.
\label{eq:Baxion}
\end{align}
This is because the dynamics following the Euler-Lagrange equation preserves the total energy of the system, and $\bm B_\textrm{Axion}$ rotates the spin while keeping the axion energy $-\mathcal{L}_\textrm{Axion}$ unchanged.
Figures~\ref{fig_analytical_calculation}(c) and \ref{fig_analytical_calculation}(d) respectively plot
the time evolution of $\bm{B}_{\mathrm{Axion}}$ and $\bm{B}_{\mathrm{Fermion}}$ for the spin dynamics shown in Fig.~\ref{fig_analytical_calculation}(a).
At $t=0$, $\bm{B}_{\text{Fermion}}$ vanishes as it should be.
On the other hand, $\bm {B}_\textrm{Axion}$ has small but nonzero $x$ and $z$ components at $t=0$ and triggers the initial spin dynamics.
The change in spin direction due to $\bm B_\textrm{Axion}$ increases the fermion energy
$E_\textrm{ss}$, which in turn generates $\bm B_\textrm{Fermion}$, resulting in a reduction of the axion energy.

\subsection{Axion-driven fermion transport}

In addition to the spin rotation, the axion carries the fermions. 
The Lagrangian density \eqref{Lagrangean_all} provides the fermion density
\begin{equation}
n_{\rm F}(x)=\pdv{\mathcal{L}}{\mu}
=1 -\frac{1}{2\pi}\partial_{x}\theta .
\label{nfermion}
\end{equation}
The total derivative in Eq.~\eqref{nfermion} indicates that the fermion number is determined only by $\theta(x)$ at the boundary. 
Indeed, the number of fermions in the region of $[0,X]$ reads
\begin{equation}
N_{X}=\int_{0}^{X}\dd{x} n_{\rm F}(x)=X-\frac{1}{2\pi}\qty[\theta(X)-\theta(0)] .
\end{equation} 
This is analogous to the low-energy ground state of quark matter governed by quantum chromodynamics at a finite baryon chemical potential under a magnetic field in which sine-Gordon solitons stabilized by chiral anomaly carry a baryon number \cite{Son_PRD2008, Brauner_JHEP2017}. 
Our axion field $\theta(x)$ does not exhibit a solitonic structure but is still responsible for fermion transport. 
The fermion current flowing from the region $[0,X]$ is found to be related to the time evolution of the axion as
\begin{equation}
\dot{N}_{X}=\dv{t}\int_{0}^{X}\dd{x} n_{\rm F}(x)=-\frac{1}{2\pi}\qty[\dot{\theta}(X)-\dot{\theta}(0)] .
\label{eq:dotNx}
\end{equation}
In a finite-size system, the transport results in the accumulation of the fermions at one of the edges, which we numerically show in the next section.

\section{Many-body fermion-spin coupled dynamics}\label{Sec:num}

Although we have demonstrated that the chemical potential gradient induces spin rotation via the axion Lagrangian within the classical spin approximation, it is nontrivial whether the inverse phenomenon of the Thouless pump survives in the many-body dynamics without introducing the axion Lagrangian. 
Here, we perform numerical calculations of the fermion-spin coupled dynamics under the Hamiltonian~\eqref{hamiltonian} using the TEBD method~\cite{Vidal_PRL2003, Vidal_PRL2004, Vidal_PRL2007, Banuls_PoS2019}. 
We consider the dynamics of quantum spins while including the motional degrees of freedom of fermions instead of introducing the axion Lagrangian.

\begin{figure}[tbp]
	\centering
	\includegraphics[width=1.02\linewidth]{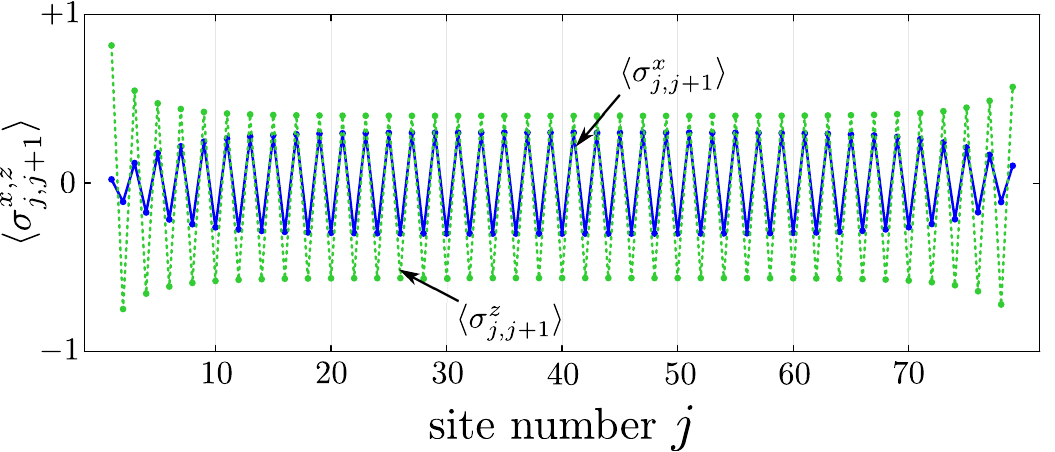}
	\caption{
    Expectation values of the link spins $\langle \sigma_{j,j+1}^{x,z}\rangle$ in the ground state for the Hamiltonian~\eqref{hamiltonian} numerically obtained by the TEBD method with imaginary-time evolution calculated with 40 fermions in $L=80$ lattice sites and 79 link spins. 
    The other parameters are chosen as $\alpha/t_h=0.5$, $\nu/t_h=1.0$, $\Delta/t_h=2/\pi$, $\beta = \nu/2$, $\mu/t_h=0.5$, and $\Delta\mu=0$.
    The staggered structure arises for both the $x$ and $z$ components. 
    The $y$ component, $\langle \sigma_{j,j+1}^y\rangle$, is zero for all $j$. 
    }
	\label{fig_spinstaggard}
\end{figure}

To this end, we first obtain the stationary state under a uniform chemical potential $\Delta\mu=0$ using the imaginary time evolution in the TEBD method. 
Figure~\ref{fig_spinstaggard} plots the expectation value of $\sigma_{j,j+1}^{x,z}$ in the ground state for $\alpha/t_h=0.5, \nu/t_h=1.0, \Delta/t_h=2/\pi$, and $\beta/t_h=0.5$.
The $y$ component $\langle \sigma_{j,j+1}^y\rangle$ is zero.
Here, we put 40 fermions in the lattice with $L=80$ sites, and $j$ takes $1$ to $79$. 
As one can see, the staggered structure appears in both the $x$ and $z$ components, which realizes the RM Hamiltonian for the fermions if we neglect the quantum nature of the link spins.
Because the values shown in Fig.~\ref{fig_spinstaggard} are the expectation values of quantum spins, their norms are not normalized.

\begin{figure}[tbp]
		\centering
            \includegraphics[width=1\linewidth]{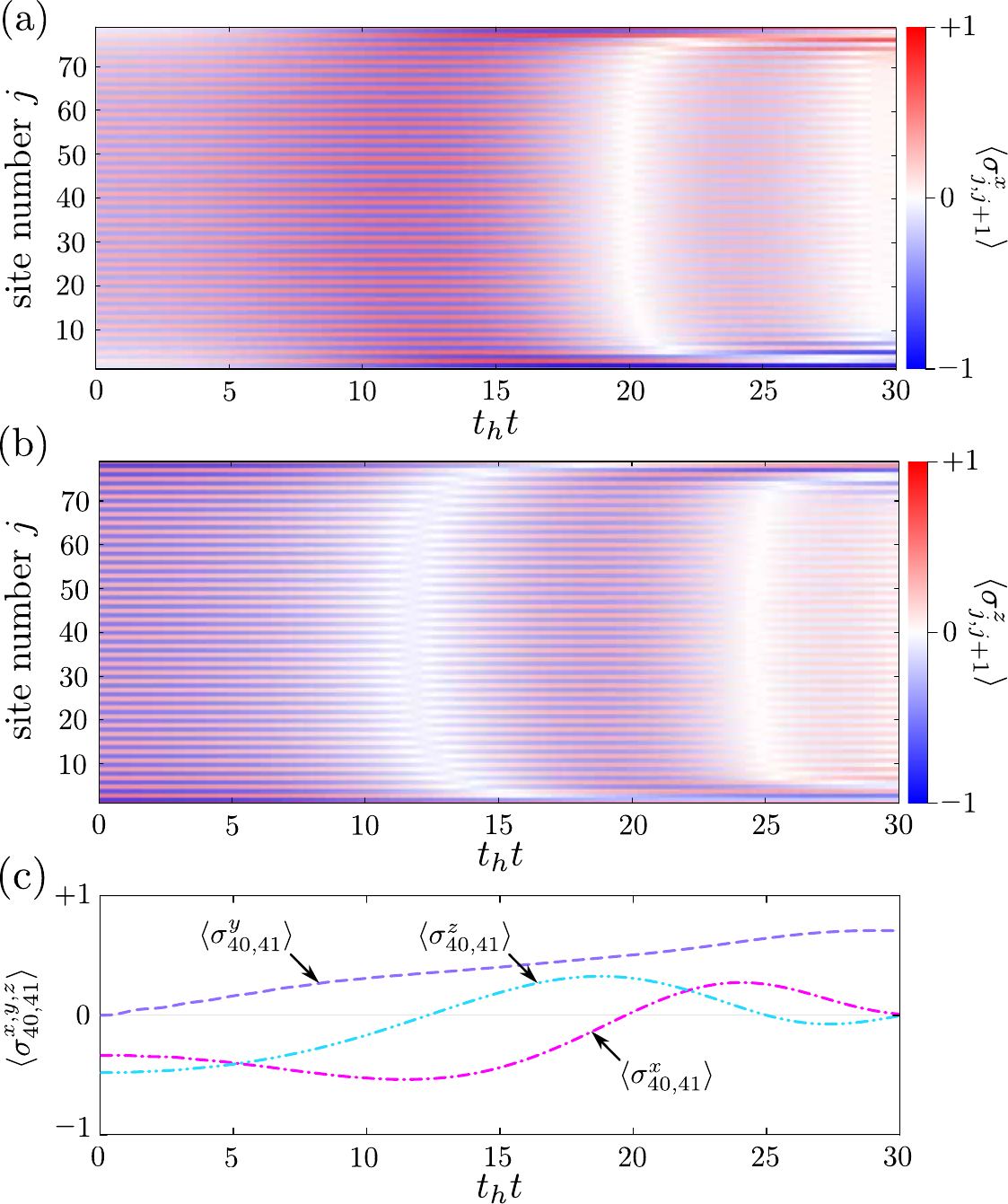}
	\caption{
		(a)(b) Time evolution of the $x$-component (a) and $z$-component (b) of the link spin expectation value, where the vertical and horizontal axes represent the site number and time, respectively. We employ the initial state shown in Fig.~\ref{fig_spinstaggard}, suddenly turn on $\Delta\mu/t_h=0.2$ at $t=0$, and solve the subsequent dynamics using the TEBD method with real-time propagation. The other parameters are the same as those for Fig.~\ref{fig_spinstaggard}.
        (c) Time evolution of the link spin expectation value at the central region, $\langle \bm\sigma_{40,41}\rangle$, where $x$ and $z$ components are the same as those in (a) and (b) at $j=40$.
    }\label{spin_dynamics_by_TEBD}
\end{figure}

Starting from the state shown in Fig.~\ref{fig_spinstaggard}, we add the chemical potential gradient $\Delta\mu=0.2t_h$ at $t=0$ and compute the subsequent time evolution. 
Figures~\ref{spin_dynamics_by_TEBD}(a) and \ref{spin_dynamics_by_TEBD}(b) show the color plots of the position (vertical axis) and time (horizontal axis) dependence of $\langle \sigma^{x}_{j,j+1}\rangle$ and $\langle \sigma^{z}_{j,j+1}\rangle$, respectively.
We also plot the time evolution of $\langle \sigma_{40,41}^{x,y,z}\rangle$ in Fig.~\ref{spin_dynamics_by_TEBD}(c).
As expected from the classical spin approximation, the $y$ component of the link spin gradually increases, and then the $x$ and $z$ components rotate about the $y$ axis.
This confirms the inverse phenomenon of Thouless pumping, i.e., the particle flow induced by the chemical potential gradient rotates the link spins.
At a later time, the $x$ component accumulates at the edge of the system, accompanied by the accumulation of fermions.

\begin{figure}[tbp]
		\centering
            \includegraphics[width=0.99\linewidth]{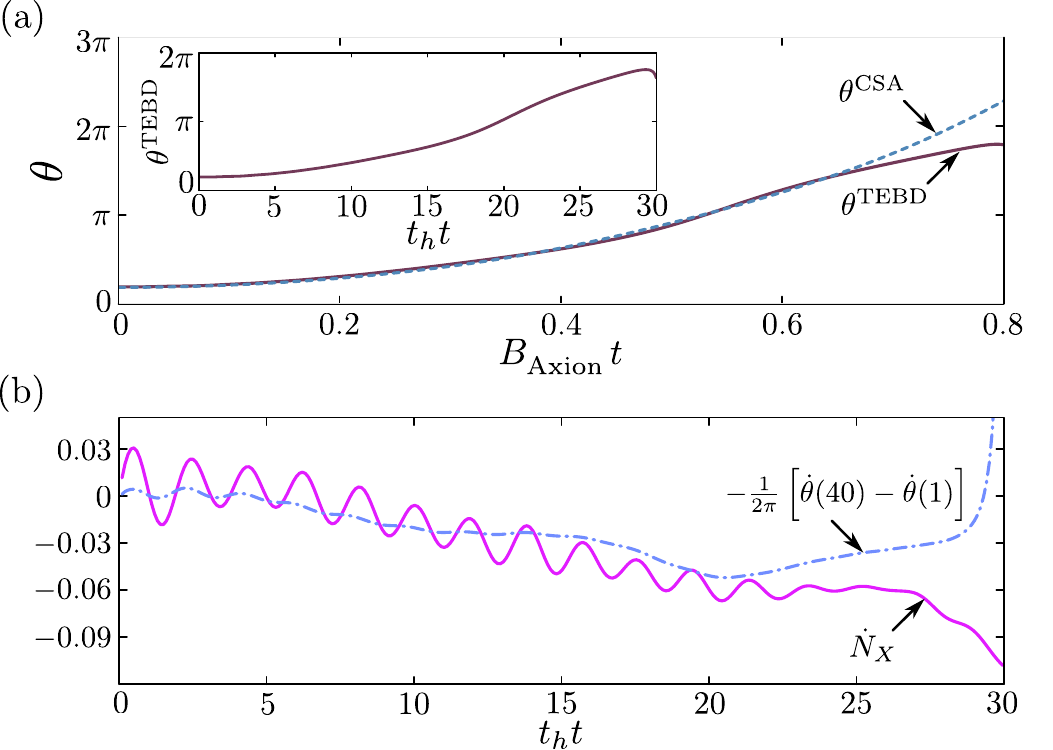}
	\caption{
                (a) Comparison between the axion field obtained by the classical spin approximation, $\theta^\textrm{CSA}$, and the one obtained from the full quantum calculation, $\theta^\textrm{TEBD}$, which coincide with each other as a function of $B_\textrm{Axion}t$ with $B_\textrm{Axion}$ obtained by each method (see main text). The data for $\theta^\textrm{CSA}$ is the same as the one shown in Fig.~\ref{fig_analytical_calculation}(b). We choose $\theta^\textrm{TEBD}$ in the central region ($j=40$) and use $B_\textrm{Axion}^\textrm{TEBD}$ at the same site. 
                Inset shows $\theta^\textrm{TEBD}(j=40)$ as a function of $t_ht$, which evolves faster than $\theta^{\rm CSA}$ in Fig.~\ref{fig_analytical_calculation}(b).
                (b) Relation between fermion transport and the axion field. 
                Each curve represents the left- and right-hand sides of Eq.~\eqref{eq:dotNx} for $X=40$ obtained from the TEBD simulation shown in Fig.~\ref{spin_dynamics_by_TEBD} by numerically taking the time derivative. 
                Although $\dot{N}_X$ exhibits small oscillations, its mean value agrees well with the prediction from the difference in the axion fields at the center and the edge of the system.
	}\label{axion_dynamics_by_TEBD}
\end{figure}

The spin dynamics shown in Fig.~\ref{spin_dynamics_by_TEBD} is well captured by the axion field theory.
In Fig.~\ref{axion_dynamics_by_TEBD}(a), we compare the time evolution of $\theta$ obtained by the full quantum (TEBD) calculation, $\theta^\textrm{TEBD}$, and the one by the classical spin approximation, $\theta^\textrm{CSA}$.
Here, we define the site-dependent $\theta^\textrm{TEBD}(j)$ by Eq.~\eqref{eq:def_axion_spin} with
replacing $s^{x,z}$ with $\langle \sigma^{x,z}_{j,j+1}\rangle$.
Note that the initial time evolution in the classical spin approximation is determined by $\bm B_\textrm{Axion}$ given in Eq.~\eqref{eq:Baxion}. 
We therefore define $B^\textrm{CSA}_\textrm{Axion}$ and $B^\textrm{TEBD}_\textrm{Axion}$ as the magnitude of $\bm B_\textrm{Axion}$ at $t=0$ for the classical spin approximation and the TEBD calculation, respectively, and plot $\theta$ as a function of $B_\textrm{Axion} t$, where, in the definition of $B_\textrm{Axion}^\textrm{TEBD}$, we carried out the same replacement of $s^{x,z}$ as in the calculation of $\theta^\textrm{TEBD}$. 
The results shown in Fig.~\ref{axion_dynamics_by_TEBD}(a) fairly agree with each other, indicating the validity of the axion field theory. 
Furthermore, by comparing the unscaled time evolution in Fig.~\ref{fig_analytical_calculation}(b) and the inset of Fig.~\ref{axion_dynamics_by_TEBD}(a), the dynamics proceeds faster for the full quantum system. 
This comes from the depletion of the spin expectation value $|\langle \bm \sigma \rangle|<1$ due to quantum coherence, which enhances $B_\textrm{Axion}$. In other words, the quantum nature of the link spins accelerates the dynamics. 

We also confirm that the above spin dynamics is linked to the fermion transport.
In Fig.~\ref{axion_dynamics_by_TEBD}(b), we compare the left- and right-hand sides of Eq.~\eqref{eq:dotNx}. 
Here, we choose $X=40$, and $N_X$ is the expectation value of the number of fermions in the left half of the system. 
It reveals that the time evolution of $\dot{N}_X$ is well captured by the difference of $\dot{\theta}$ in the central region and at the edge although $\dot{N}_{X}$ includes small oscillations. 
This transport dynamics accumulates fermions around the edge at $j=80$, which accompanies the accumulation of $\sigma^x$ as seen in Fig.~\ref{spin_dynamics_by_TEBD}(a) via the spin-dependent on-site potential. Spin accumulation destroys the staggered spin structure in the central region, and hence the system is no longer described by the classical spin approximation with the axion Lagrangian. 

\section{Proposal for Experimental Implementation}\label{Secexp}
\begin{figure}
    \centering
    \includegraphics[width=0.95\linewidth]{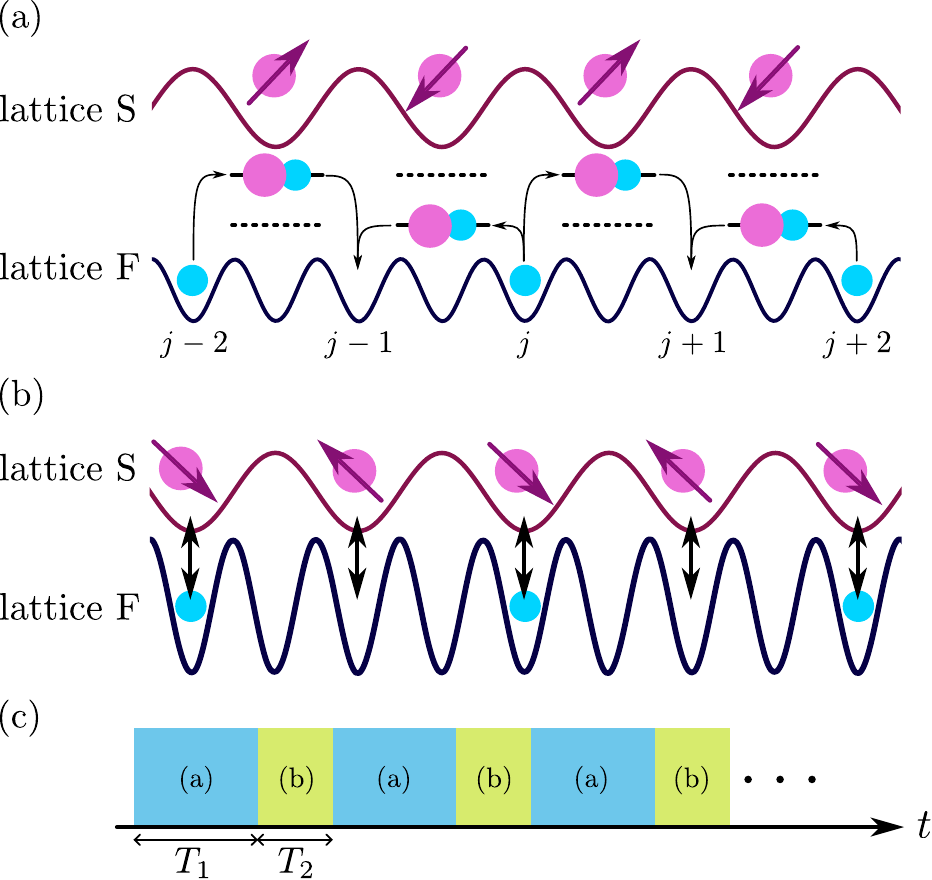}
    \caption{Schematic of the experimental setup. (a) $\mathbb{Z}_2$ gauge field is implemented by using two 1D optical lattices; Lattice S for two-level atoms, which work as spin-1/2 link spins, and Lattice F for spinless fermions, where the lattice constant for the former lattice is twice that of the latter~\cite{Titas_SciPostPhys2022}. When Lattice S is filled at filling fraction 1, the interaction between the atoms in two lattices prevents the spinless atoms from occupying the overlapping sites. According to the second-order perturbation, the spin-dependent interaction between the link spins and fermions in the intermediate state realizes the spin-dependent hopping for fermions.
    (b) To realize the $\sigma^x$ dependent on-site potential, we periodically apply a $\pm\pi/2$ pulse and simultaneously shift and back Lattice S by half of its period. 
    In the intermediate state, the $x$ and $z$ components of the link spins are swapped by the $\pi/2$ pulse, and the link spin sites are overlapped with fermion sites, inducing the $\sigma^x$ dependent on-site interaction for fermions. We deepen Lattice F to prevent the fermion hopping during this configuration.
    (c) Time sequence for switching the configurations (a) and (b). 
    By tuning the period $T_1$ and $T_2$, we can control the parameters in the Hamiltonian.
    }
    \label{experimental_setup}
\end{figure}

Finally, we comment on the experimental setup to realize the Hamiltonian~\eqref{hamiltonian}. 
The method for implementing the Hamiltonian $\hat{\mathcal H}'$ including the $\mathbb{Z}_2$ gauge field was proposed in Ref.~\cite{Titas_SciPostPhys2022}. 
We prepare two one-dimensional lattices, S and F,
with the lattice constant of S being twice that of F.
Lattices S and F are for the spins and the fermions, respectively.
We load atoms with two internal states ($\ket{\uparrow}, \ket{\downarrow}$) and spinless fermions into lattices S and F, respectively.
The depth of lattice S is set to be deep enough so that the atoms in lattice S are in the Mott insulating state with filling fraction 1.
On the other hand, due to the repulsive interaction with the atoms in lattice S, the fermions in lattice F occupy lattice sites that do not overlap with the sites in lattice S, which are labeled $\cdots, j-1, j, j+1, \cdots$ in Fig.~\ref{experimental_setup}(a).
The hopping integral between the next nearest neighboring sites in lattice F is evaluated with the second-order perturbation. In the intermediate state of the hopping, where the fermions are on the sites of overlap with the spins, the fermion-spin interaction energy depends on the $z$ component of the link spins, which gives the $\sigma^z$-dependent hopping integral.

To introduce the second term of the Hamiltonian~\eqref{hamiltonian}, i.e., the on-site potential dependent on the $x$ component of the neighboring link spin, we propose to use Floquet engineering techniques \cite{Jotzu_nature2014,schweizer_nphys2019}. 
We repeat to shift and back lattice S by half of the lattice constant and simultaneously apply the $\pm\pi/2$ pulse to the atoms in lattice S.
In the intermediate state shown in Fig.~\ref{experimental_setup}(b), the link spins and the fermions are in the same spatial position. Thus, the fermions undergo an interaction dependent on the $z$ component of the spin, which is the $x$ component of the link spin in Fig.~\ref{experimental_setup}(a).
We also deepen lattice F to suppress the hopping in the configuration shown in Fig.~\ref{experimental_setup}(b).
We switch the two configurations of Figs.~\ref{experimental_setup}(a) and \ref{experimental_setup}(b) with durations $T_1$ and $T_2$ repeatedly, as illustrated in Fig.~\ref{experimental_setup}(c).
When we choose $T_1$ and $T_2$ small enough such that the corresponding frequency is much larger than the energy scales of the system, the time-averaged Hamiltonian corresponds to Eq.~\eqref{hamiltonian}.


\section{Conclusion}\label{Secsummary}

In this paper, we investigated the fermion-spin coupled dynamics in a one-dimensional lattice. 
Starting from the Hamiltonian for the spinless fermions under the $\mathbb{Z}_2$ dynamical gauge field, where the Ising spins on the links behave as the gauge field, we introduced the on-site potential for the fermions dependent on the $x$ component of the link spins. 
Under a properly chosen set of parameters at half-filling of fermions, the link spins exhibit a staggered structure in both the $x$ and $z$ components, realizing the Rice-Mele Hamiltonian for the fermions. 
The Rice-Mele Hamiltonian is transformed to that of the normal insulator under the chiral gauge transformation, which accompanies the additional coupling between the axion field $\theta$ and the electric field $E$ as $\theta E$. 
In our model, the direction of the link spins plays the role of the axion field $\theta$, and hence it evolves in time. 
Within the classical spin approximation, the spin dynamics following the Euler-Lagrange equation predicts the rotation of link spins in response to the applied chemical potential gradient. 
We can interpret this dynamics as the inverse phenomenon of Thouless pumping. 
We note that the axion Lagrangian does not directly induce the dynamics of the axion field because of energy conservation. 
The axion Lagrangian rotates the spins in the direction in which the axion field does not change. This rotation increases the total energy of the occupied fermion band, which enables the axion field to evolve so that the energy corresponding to the axion Lagrangian decreases while conserving the total energy of the system. 

To confirm our prediction, we also solved the fermion-spin coupled many-body dynamics by utilizing the TEBD method. 
Remarkably, we found that our numerical results fairly agree with those using the classical spin approximation as a function of time scaled by the magnitude of the effective magnetic field $B_\textrm{Axion}$ originating from the axion Lagrangian.
This agreement clearly indicates the validity of the axion field theory.
On the other hand, since the spin correlation suppresses the norm of the spin expectation value, $B_\textrm{Axion}$ becomes larger for the full-quantum calculation. 
As a result, the full-quantum dynamics proceeds faster than the classical spin dynamics compared in the same time scale.
Furthermore, the evolution of the axion field is accompanied by the transport of fermions. 
From the axion Lagrangian, it follows that the time derivative of the fermion numbers in a certain region is given by the difference between the time derivatives of the axion fields at the boundaries [Eq.~\eqref{eq:dotNx}].
Particularly in a finite-size system, the axion field nonuniformly evolves and induces fermion flow inside the system, leading to the accumulation of fermions on one side.
We have numerically verified that Eq.~\eqref{eq:dotNx} qualitatively holds during the many-body quantum dynamics.
We also proposed possible experimental implementations of our model using cold atoms. 

There are several remaining issues. Because we considered the energy-conserving unitary dynamics in this study, particles and spins accumulated on one side over time under the chemical potential gradient, resulting in the staggered spin structure to disappear. Under a temporally varying potential gradient, which corresponds to an AC electric field, it may be possible to maintain the staggered spin structure for a longer period. 
Furthermore, depending on the AC frequency, magnetic resonance in the spin system may be induced~\cite{Sekine_PRB2016}. To observe stable resonance phenomena, it may be better to make the system open and allow energy dissipation. 
On the other hand, if energy is injected appropriately, we may investigate particle generation. It is also interesting to examine the case where the axion field varies spatially, such as in the presence of domain walls. According to Eq.~\eqref{nfermion}, when the axion field jumps by $\pi$ at the domain wall, a fractionalized fermion appears there~\cite{Gonzalez_PRL2020}. Investigating how such domain walls, accompanied by fractional particles, respond to an external field is an intriguing issue.

\acknowledgments
We thank K. Fujimoto for the fruitful discussions.
This work was supported by JSPS KAKENHI (Grant Numbers JP24K00557 and JP24K22858). 
K.F. was also supported by Maki Makoto Foundation. 


\appendix
\begin{appendix}
\section{Energy comparison for possible staggered spin configurations}\label{energy_minimums}
For spins to create a two-site sublattice structure, there are three possibilities: (I) a staggard spin state in the $z$ direction with a uniform $x$ component, (II) a staggard spin state both in the $x$ and $z$ directions, and (III) a staggard spin state only in the $x$ direction with a uniform $z$ component. For simplicity, we assume that the directions of spins on sublattices $\rm A$ and $\rm B$ are parametrized as follows.
\begin{description}
    \item[case I]
    \begin{subequations}
    \begin{align}
        (\sigma^x_\mathrm{A}, \sigma^y_\mathrm{A}, \sigma^z_\mathrm{A})&=(\sin\vartheta\cos\varphi, \sin\vartheta\sin\varphi, \cos\vartheta)\\
        (\sigma^x_\mathrm{B}, \sigma^y_\mathrm{B}, \sigma^z_\mathrm{B})&=(\sin\vartheta\cos\varphi, \sin\vartheta\sin\varphi, -\cos\vartheta)
    \end{align}
    \label{eq:caseI}
    \end{subequations}
    \item[case I\hspace{-1.2pt}I]
    \begin{subequations}
    \begin{align}
        (\sigma^x_\mathrm{A}, \sigma^y_\mathrm{A}, \sigma^z_\mathrm{A})&=(\sin\vartheta\cos\varphi, \sin\vartheta\sin\varphi, \cos\vartheta)\\
        (\sigma^x_\mathrm{B}, \sigma^y_\mathrm{B}, \sigma^z_\mathrm{B})&=(-\sin\vartheta\cos\varphi, \sin\vartheta\sin\varphi, -\cos\vartheta)
    \end{align}
    \end{subequations}
    \item[case I\hspace{-1.2pt}I\hspace{-1.2pt}I]
    \begin{subequations}
    \begin{align}
        (\sigma^x_\mathrm{A}, \sigma^y_\mathrm{A}, \sigma^z_\mathrm{A})&=(\sin\vartheta\cos\varphi, \sin\vartheta\sin\varphi, \cos\vartheta)\\
        (\sigma^x_\mathrm{B}, \sigma^y_\mathrm{B}, \sigma^z_\mathrm{B})&=(-\sin\vartheta\cos\varphi, \sin\vartheta\sin\varphi, \cos\vartheta)
    \end{align}
    \label{eq:caseIII}
    \end{subequations}
\end{description}
The case we consider in the main text corresponds to case (I\hspace{-1.2pt}I).
We substitute the above spin configurations into Eq.~\eqref{eq:fermion_dispersion}, integrate the lower band in the first Brillouin zone, and obtain the $(\vartheta,\varphi)$ dependence of the energies as 
\begin{widetext}
    \begin{align}
        \frac{2E_{\text{I}}(\vartheta, \varphi)}{L}
        &=-\mu+(2\beta-\nu)\sin\vartheta\cos\varphi
        -\frac{4t_h}{\pi}{\mathfrak E}\left(1-\frac{\alpha^2}{t_h^2}\cos^2\vartheta\right),\label{Energy_1}\\
        \frac{2E_{\text{I\hspace{-1.2pt}I}}(\vartheta, \varphi)}{L}
        &=-\mu-\frac{4t_h}{\pi}\sqrt{1+\frac{\nu^2}{4t_h^2}\sin^2\vartheta\cos^2\varphi}\,{\mathfrak E}\left(\frac{1-\frac{\alpha^2}{t_h^2}\cos^2\vartheta}{1+\frac{\nu^2}{4t_h^2}\sin^2\vartheta\cos^2\varphi}\right),\label{Energy_2}\\
        \frac{2E_{\text{I\hspace{-1.2pt}I\hspace{-1.2pt}I}}(\vartheta, \varphi)}{L}
        &=-\mu+\Delta\cos\vartheta-\frac{4(t_h+\alpha\cos\vartheta)}{\pi}\sqrt{1+\frac{\nu^2\sin^2\vartheta\cos^2\varphi}{4(t_h+\alpha\cos\vartheta)^2}}{\mathfrak E}\left(\frac{1}{
        1+\frac{\nu^2\sin^2\vartheta\cos^2\varphi}{4(t_h+\alpha\cos\vartheta)^2}}\right),\label{Energy_3}
    \end{align}
\end{widetext}
where, $E_{\text{I\hspace{-1.2pt}I}}$ is identical to $E_\textrm{ss}$ in Eq.~\eqref{Energy}.

We first compare $E_{\text{I\hspace{-1.2pt}I}}(\vartheta,\varphi)$ and $E_{\text{I}}(\vartheta,\varphi)$.
Note that $2\beta-\nu$ in Eq.~\eqref{Energy_1} works as an effective Zeeman field along the $x$ direction. When this term gives a nonzero contribution,
the system tends to have a net spin along the $x$ direction, and case I is favored. 
We therefore choose $\beta=\nu/2$, at which Eq.~\eqref{Energy_1} becomes independent of $\nu$ and $\varphi$. 
In addition, because ${\mathfrak E}(\kappa)$ is a monotonically decreasing function of $\kappa$,
$2E_{\text{I}}(\vartheta,\varphi)/L$ takes the minimum value $2E_\textrm{I}^\textrm{min}/L=-\mu-(4t_h/\pi) E(1-\alpha^2/t_h^2)$
at $\vartheta=0$.
Since the spin configurations in cases I and I\hspace{-1.2pt}I are identical when $\vartheta=0$, we conclude that $E_{\text{I\hspace{-1.2pt}I}}(\vartheta,\varphi)$ has the same or a lower minimum energy than $E_{\text{I}}(\vartheta,\varphi)$ when $\beta=\nu/2$.

\begin{figure}[t]
		\centering
            \includegraphics[width=0.95\linewidth]{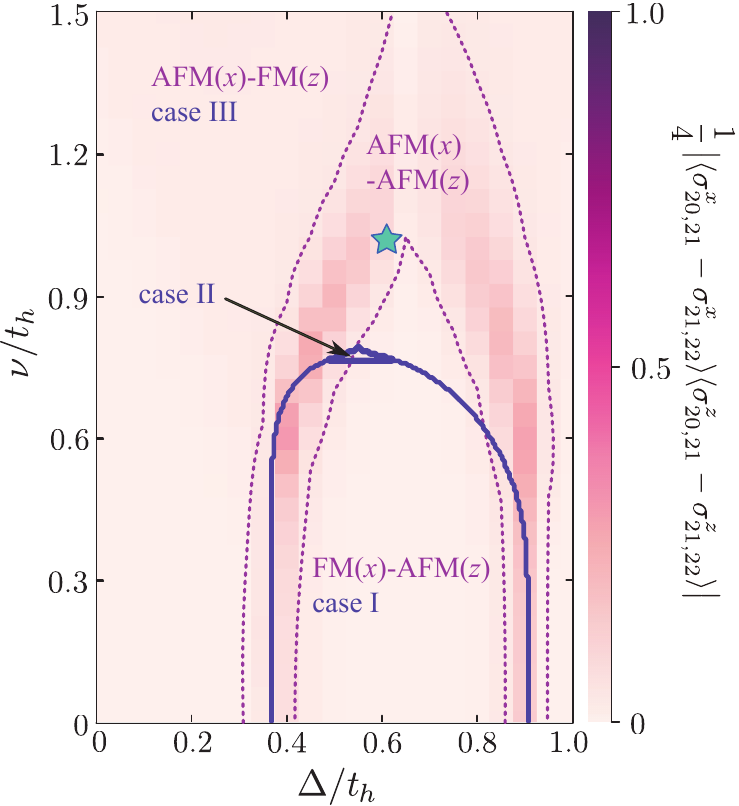}
        \caption{
        Ground-state phase diagram at $\alpha/t_h=0.5$ and $\beta=\nu/2$. 
        The solid violet curves are the phase boundaries wihtin the classical spin approximation determined by Eqs.~\eqref{eq:caseI}-\eqref{eq:caseIII}.
        The heat map shows the differences in the vertical and horizontal spin components $|\langle\sigma^x_{20,21} - \sigma^x_{21,22}\rangle\langle\sigma^z_{20,21} - \sigma^z_{21,22}\rangle|/4$ in the ground state calculated by the TEBD method with imaginary-time evolution for $L=40$. 
        The dashed magenta curves are the contour at $0.1$, being the guide for the eyes of the phase boundaries. 
        The labels FM/AFM($x$) and FM/AFM($z$) in each region stand for the ferromagnetic/antiferromagnetic orders in the spin components $x$ and $z$, respectively.
        The star indicates the parameters for the numerical calculation used in Sec.~\ref{Sec:num}.
	}\label{spin_ground}
\end{figure}

Next, we compare $E_{\text{I\hspace{-1.2pt}I}}(\vartheta,\varphi)$ and $E_{\text{I\hspace{-1.2pt}I\hspace{-1.2pt}I}}(\vartheta,\varphi)$.
From the $\kappa$ dependence of ${\mathfrak E}(\kappa)$, Eqs.~\eqref{Energy_2} and \eqref{Energy_3} are minimized with respect to $\varphi$ at $\varphi=0$. 
From the fact that $E_{\text{I\hspace{-1.2pt}I}}(\pi/2,0)=E_{\text{I\hspace{-1.2pt}I\hspace{-1.2pt}I}}(\pi/2,0)$,
$E_{\text{I\hspace{-1.2pt}I}}(\vartheta,\varphi)$ has a smaller minimum than $E_{\text{I\hspace{-1.2pt}I\hspace{-1.2pt}I}}(\vartheta,\varphi)$, at least if $E_{\text{I\hspace{-1.2pt}I\hspace{-1.2pt}I}}(\vartheta,0)$ has a minimum at $\vartheta=\pi/2$.
Since the last term on the right-hand side of Eq.~\eqref{Energy_3} is not symmetric with respect to $\vartheta=\pi/2$, this condition is satisfied by tuning the value of $\Delta$ so that $E_{\text{I\hspace{-1.2pt}I\hspace{-1.2pt}I}}(\vartheta,0)$ becomes symmetric with respect to $\vartheta=\pi/2$ around $\vartheta\simeq\pi/2$.
For further comparison, we need numerical calculations.

We also have to verify that the obtained spin-staggered state I\hspace{-1.2pt}I has a lower energy than a uniform spin configuration by properly choosing the value of $\Delta$.
When we have a uniform spin configuration tilted in the $x$-$z$ plane, $(\sigma^x, \sigma^y, \sigma^z)=(-\sin\vartheta, 0, -\cos\vartheta)$, the fermionic Hamiltonian is given by
\begin{align}
    \hat{\mathcal{H}}_\textrm{uni}=&
    -\frac{L}{2}\Delta\cos\vartheta-L \beta\sin\vartheta -(\mu-\nu\sin\vartheta)\sum_j \hat{a}_j^\dagger \hat{a}_j\nonumber\\
    &-(t-\alpha\cos\vartheta)\sum_{j=1}^{L-1}\left(\hat{a}^{\dagger}_j\hat{a}_{j+1}+\text{h.c.}\right)\label{energy_F},
\end{align}
from which we obtain the system's energy per two sites at half-filling as
\begin{align}
    \frac{2E_\textrm{uni}(\vartheta)}{L}=&-\mu-\Delta\cos\vartheta-(2\beta-\nu)\sin\vartheta\nonumber \\
    &-\frac{4(t_h-\alpha\cos\vartheta)}{\pi} \label{Energy_uni}\\
    \ge& -\mu-\frac{4t_h}{\pi} - \sqrt{\left(\Delta-\frac{4\alpha}{\pi}\right)^2+(2\beta-\nu)^2}.
\end{align}
Because $2E_\textrm{I\hspace{-1.2pt}I}(\vartheta,0)/L\le -\mu-4t_h/\pi$, the energy of the uniform spin configuration becomes larger than that of case I\hspace{-1.2pt}I
when $\Delta\simeq 4\alpha/\pi$ and $\beta\simeq \nu/2$. 

Figure \ref{spin_ground} shows the phase diagram in the $(\Delta,\nu)$ space obtained in the classical spin approximation by comparing the minimum energies of Eqs.~\eqref{Energy_1}-\eqref{Energy_3}. 
The phase boundary is depicted with the solid violet curves. 
Here, we fix $\beta=\nu/2$ and $\alpha=t_h/2$. 
Although case I with $\vartheta=\pi/2$ or case I\hspace{-1.2pt}I\hspace{-1.2pt}I with $\vartheta=0$ or $\pi$ gives uniform spin configurations, we do not specify these cases from other $\vartheta$ and label ``case I'' or ``case I\hspace{-1.2pt}I\hspace{-1.2pt}I''.
One can see that case I\hspace{-1.2pt}I arises in the small region in the parameter space. 
In Fig.~\ref{spin_ground}, we also plot the heat map of $\frac{1}{4}|\langle\sigma^x_{20,21} - \sigma^x_{21,22}\rangle\langle\sigma^z_{20,21} - \sigma^z_{21,22}\rangle|$, which becomes nonzero when antiferromagnetic order arises in both the spin components $x$ and $z$, in the ground state for $L=40$ obtained by the TEBD method with imaginary time evolution. 
The dashed magenta curves represent the contour at 0.1, serving as a guide for the magnetic phase boundary. The labels FM/AFM($x$) and FM/AFM($z$) in each region represent the ferromagnetic/antiferromagnetic orders in the spin components $x$ and $z$, respectively. 
The TEBD result qualitatively agrees with the classical spin approximation, but the region of AFM($x$)-AFM($z$) is largely expanded from that of case I\hspace{-1.2pt}I. We use the parameters marked with a star for the numerical calculation used in Sec.~\ref{Sec:num}.

\end{appendix}

\nocite{*}

\bibliography{reference.bib}

\end{document}